\documentclass[aps,prb, notitlepage, preprintnumbers, nofootinbib, twocolumn, 
amsmath,amssymb,superscriptaddress,floatfix,scrartcl, 
]{revtex4-2}

\usepackage{amssymb}
\usepackage{color}
\usepackage{graphicx}
\usepackage{mathrsfs}
\usepackage[unicode=true,pdfusetitle,bookmarks=false,colorlinks=true,citecolor=blue,urlcolor=blue,linkcolor=red]{hyperref}
\usepackage{bm}
\usepackage{braket}
\usepackage{tikz}
\usepackage{comment}
\usepackage[normalem]{ulem} 

\usetikzlibrary{decorations.markings, calc, arrows.meta, bending}

\newcommand{\average}[1]{[#1]}

\usepackage{cleveref}
\crefname{appendix}{App.}{Apps.}
\crefname{equation}{Eq.}{Eqs.}
\crefname{figure}{Fig.}{Figs.}
\crefname{table}{Tab.}{Tabs.}
\crefname{section}{Sec.}{Secs.}

\begin{document}

\title{Tripartite entanglement in chaotic eigenstates}

\author{Junjia Zhang}
\affiliation{%
Department of Physics, Princeton University, Princeton, New Jersey 08544, USA%
}%

\author{Ramanjit Sohal}
\affiliation{
Pritzker School of Molecular Engineering, University of Chicago, Chicago, IL 60637, USA}

\author{Shinsei Ryu}
\affiliation{%
Department of Physics, Princeton University, Princeton, New Jersey 08544, USA%
}%

\begin{abstract}
It is now well-established that classical statistical mechanics emerges from the entanglement structure of quantum chaotic systems, as quantified by the (subsystem) eigenstate thermalization hypothesis (ETH). While this statement rests on the well-studied \emph{bipartite} entanglement of the eigenstates of such systems, recent years have shown that many-body states are often characterized by multipartite entanglement, raising the question: what are the \emph{universal} multipartite entanglement features of chaotic eigenstates?
We answer this question in part by studying a tripartite generalization of the ergodic bipartition ansatz through the lens of two tripartite entanglement probes, the R\'enyi tripartite multi-entropy and the R\'enyi reflected entropy. We derive leading-order analytical expressions for the ensemble averages of both quantities using a saddle-point analysis. 
Beyond these measures, we obtain the average of the R\'enyi-2 reflected density matrix, $\rho_{AA^*}^{(2)}$,
which controls the R\'enyi-2 reflected entropy. 
We show that $\rho_{AA^*}^{(2)}$ in a chaotic eigenstate is well-described by this ensemble average
which,
unlike an ordinary reduced density matrix,
takes three qualitatively distinct forms across subsystem-size regimes, which may be viewed as a tripartite generalization of subsystem ETH.
We benchmark our analytical predictions against exact diagonalization of the mixed-field Ising model.

\end{abstract}

\maketitle

\section{Introduction}

A fundamental question across a range of fields in physics is how statistical mechanics emerges from the dynamics of an isolated quantum system. It is now understood that a large class of many-body systems, known as non-integrable or \emph{quantum chaotic} systems, can act as their own baths, leading to effectively thermal behavior \cite{Deutsch_1991, Srednicki_1994, Srednicki_1999, Rigol_2008}.
This is formalized with the eigenstate thermalization hypothesis (ETH), which makes the claim that local correlators in high-temperature eigenstates of such systems, which we call \emph{chaotic eigenstates}, look thermal.
A stronger claim is made by subsystem ETH \cite{Dymarsky_2018}, which asserts that the reduced density matrix in such an eigenstate is close to a thermal density matrix. It is in this sense that thermal entropy emerges from the entanglement of chaotic systems.
As probes of this entanglement,
von Neumann and R\'enyi bipartite entanglement entropies have been examined to explore the structure of these chaotic eigenstates \cite{Garrison_2018, Lu_2017, Vidmar_2017, Murthy_2019, Bianchi_2022, Rodriguez_Nieva_2024, Langlett_2025}.
In particular, an analytical understanding of the \emph{universal} properties of these quantities has been obtained using the ``ergodic bipartition" ansatz.

Specifically, it has been conjectured \cite{Deutsch_2010, Lu_2017, Murthy_2019} that in a system partitioned into two subsystems $A$ and $B$, a chaotic eigenstate takes the following form:
\begin{equation}
    |E\rangle = \sum_{i, j} M_{ij} |E_i^A\rangle_A \otimes |E_j^B\rangle_B.
    \label{ergodic_bipartition}
\end{equation}
Here, $|E_i^A\rangle$ and $|E_j^B\rangle$ are energy eigenstates of subsystems $A$ and $B$ with energies $E_i^A$ and $E_j^B$. The coefficient $M_{ij}$ is described as \cite{Murthy_2019}:
\begin{equation}
    M_{ij} = e^{-S(E_i^A + E_j^B)/2} F(E_i^A + E_j^B - E)^{1/2} C_{ij},
    \label{ergodic_bipartition_coefficient}
\end{equation}
where $S(E)$ is the microcanonical entropy (i.e., the logarithm of the density of states at energy $E$), $F(\Delta)$ is a window function sharply peaked at $\Delta = 0$, and $C_{ij}$ is a Gaussian random variable satisfying 
$[C_{ij}] = 0$ and $[C_{ij}C_{i'j'}] = \delta_{ii'}\delta_{jj'}$. 
Here, and henceforth, the square bracket $[\cdots]$ represents the ensemble average.
The ansatz \eqref{ergodic_bipartition} is of the form suggested by ``canonical typicality" \cite{Goldstein_2006, Popescu_2006}, and is consistent with subsystem 
ETH. 
The ergodic bipartition ansatz can be interpreted as stating that chaotic eigenstates mimic those drawn from a Haar random ensemble subject to energy conservation.
We refer to such 
states as (energy) constrained Haar random states.

Using this ansatz, Ref.~\cite{Lu_2017} derived analytical expressions for bipartite R\'enyi entropies of chaotic eigenstates, and showed that they agree well with exact diagonalization results, 
supporting the picture that chaotic eigenstates share 
statistical properties with constrained Haar random states. Moreover, 
the ergodic bipartition ansatz allowed further understanding of the structure of chaotic eigenstates. 
For example, it predicts that R\'enyi entropies are convex in subsystem size when the R\'enyi index $n > 1$, distinguishing chaotic eigenstates from thermal states or canonical thermal pure quantum (CTPQ) states \cite{Nakagawa_2018} which instead predict constant volume-law coefficient \cite{Lu_2017}. When $n=1$, on the other hand, Ref.~\cite{Murthy_2019} derived, using the ergodic bipartition ansatz, that the von Neumann entropy exhibits a universal correction proportional to the square root of the system size when the two subsystems have nearly equal volumes.
At the same time, chaotic eigenstates also exhibit structures beyond those of constrained Haar random states. For example, recent work \cite{Rodriguez_Nieva_2024} showed that higher moments of the von Neumann entropy in ensembles of chaotic eigenstates can deviate from the constrained Haar random ensemble, even when the average entropies of the two ensembles match well.

This body of work rests almost entirely on bipartite entanglement. 
However, quantum mechanical systems can exhibit multipartite entanglement to which bipartite probes are insensitive. Indeed, recent years have shown that quantum phases of matter in fact often can be characterized by their multipartite entanglement structure. 
It is thus natural to ask whether such multipartite correlations are also features of quantum chaotic systems.
The purpose of this work is to extend this paradigm to chaotic eigenstates and probe the \emph{universal} features of this richer entanglement structure in these states.

Specifically, 
we make use of two entanglement measures that have been developed in recent years specifically to capture tripartite correlations: the multi-entropy \cite{Gadde:2022cqi, Penington2023, Gadde2023, Harper:2024ker} and the reflected entropy \cite{Dutta2021}. Both probe correlations beyond simple Bell-type bipartite entanglement.
Indeed, the reflected entropy has been shown to probe tripartite entanglement beyond that of pure bipartite or Greenberger-Horne-Zeilinger (GHZ) types of entanglement \cite{Zou:2020bly, Siva_2022}.
The multi-entropy, on the other hand, is a multipartite generalization of the (R\'enyi) bipartite entanglement entropy that is symmetric among all subsystems.
Specifically, it was found that, similar to bipartite entanglement entropy which is determined by minimal bipartitions of holographic states and random tensor networks \cite{Ryu_2006, Hayden2016}, the tripartite R\'enyi-2 multi-entropy in these systems is determined by minimal tripartitions \cite{Penington2023}.

Though initially proposed in holographic settings, both quantities have been studied in a wide range of physical systems, including but not limited to (conformal) field theories 
\cite{Bueno2020, Bueno2020_2, Kusuki2020, Moosa2020, Kudler-Flam2021, Berthiere_2023, Gadde:2022cqi, Gadde2023, Harper:2024ker}, tensor networks \cite{Akers2022, Akers2023, Akers2024}, and topological phases of matter \cite{Siva_2022, Liu_2022, Liu_2023, Sohal_2023, Berthiere_2021, Yin_2023, Liu_2025, Yuan:2025dgx, Sohal_2026}. These tripartite quantities can capture features that are not encoded in bipartite entanglement entropy. For example, in topological phases of matter, it has been found that the reflected entropy 
can diagnose obstructions to fully gapping edge modes \cite{Siva_2022, Liu_2022, Liu_2023}.
Equipped with these finer probes, the natural question is the extent to which the constrained Haar random state description of chaotic eigenstates,
successful in the bipartite setting,
carries over to the tripartite setting. We take up this question in the remainder of the paper.

Specifically, for a quantum chaotic system tripartitioned into subsystems $A$, $B$ and $C$, we consider the following extension of the ergodic bipartition ansatz, 
\begin{equation}
    |E\rangle = \sum_{i, j, k} M_{ijk}|E_i^A\rangle_A |E_j^B\rangle_B |E_k^C\rangle_C,
    \label{ergodic_tripartition}
\end{equation}
where, 
\begin{equation}
    M_{ijk} = e^{-S(E_i^A+E_j^B+E_k^C)/2} F(E_i^A+E_j^B+E_k^C-E)^{1/2} C_{ijk}.
    \label{ergodic_tripartition_coefficient}
\end{equation}
Here, $F(E_i^A+E_j^B+E_k^C-E)$ is a window function sharply peaked around $E_i^A + E_j^B + E_k^C = E$. $C_{ijk}$ is a Gaussian random variable satisfying 
$[C_{ijk}] = 0$ and 
$[C_{ijk}C_{i'j'k'}] = \delta_{ii'}\delta_{jj'}\delta_{kk'}$. 
We note that Eq.\ \eqref{ergodic_tripartition} has been previously employed to calculate entanglement negativity \cite{Lu_2020} and entanglement negativity transitions \cite{McBride_2023} in chaotic eigenstates. 
In these works the validity of the ansatz was assumed and numerical checks focused on qualitative scaling behavior rather than quantitative agreement.

In our work, based on Eq.\ \eqref{ergodic_tripartition}, we obtain analytical expressions for R\'enyi multi-entropy and R\'enyi reflected entropy, 
to probe tripartite correlations 
that are not captured by 
bipartite entanglement entropy or entanglement negativity. 
We also study the analytical form of the average of the reduced density matrix $\rho_{AA^*}^{(2)}$, which is involved in the definition of $(2, n)$-R\'enyi reflected entropy and operator entanglement.
We test these predictions by exact diagonalization of the mixed-field Ising model and find good agreement: the constrained-Haar ensemble reproduces both the average R\'enyi-2 multi-entropy and the full average $\rho_{AA^*}^{(2)}$ at the level of trace distance. 
Moreover, the parameter regime over which the R\'enyi-2 multi-entropy agrees with the constrained-Haar prediction closely tracks that of the bipartite R\'enyi-2 entropy, even where genuinely tripartite correlations are present. We therefore find no evidence that tripartite correlations lead to an earlier breakdown of the constrained-Haar description at the level of ensemble averages.

Two features make the tripartite problem qualitatively richer than its well-studied bipartite counterpart, 
and they constitute the main message of this work. 
First, the bipartite entanglement of random states is controlled by two competing contributions, i.e., the two permutations of the replica computation,
so that, as the subsystem sizes are varied, a bipartite entropy passes through a single transition between two phases. 
The tripartite probes studied here, the (R\'enyi) multi-entropy and the (R\'enyi) reflected entropy, instead support additional phases that appear when the three subsystems are of comparable size, reflecting a genuinely multipartite competition among several saddle points. 
Imposing energy conservation modifies the analytic expressions for these probes.
In particular, they no longer take the simple cycle-counting form of the unconstrained case. However, this does not alter this leading-order phase structure: the energy constraint changes the value of each probe within a phase, and its subleading finite-size corrections, rather than creating or removing phases.

Second, this richer structure is already visible at the level of the reduced density matrix. 
We obtain the ensemble average of the R\'enyi-2 reflected density matrix $\rho_{AA^*}^{(2)}$, the spectrum of which controls the $(2,n)$-R\'enyi reflected entropies and the operator entanglement of $\rho_{AB}$,
and find that it exhibits qualitatively distinct behavior in different subsystem-size regimes, in contrast to the bipartite reduced density matrix, 
which takes a single thermal form. 

The remainder of this paper is organized as follows.
In Section~\ref{sec:quantities_of_interest}, we introduce the two probes of tripartite entanglement used throughout, the reflected entropy and the multi-entropy.
In Section~\ref{section_2}, we derive the analytical predictions for the R\'enyi multi-entropy and the R\'enyi reflected entropy from the energy-constrained Haar random ansatz [Eq.\ \eqref{ergodic_tripartition}], including their saddle point analysis and the average reduced density matrix $\average{\rho_{AA^*}^{(2)}}$ together with its three qualitative forms in different subsystem-size regimes.
In Section~\ref{sec:exact_diagonalization_multi_entropy}, we test these predictions against exact diagonalization of the mixed-field Ising model: we compare the R\'enyi-2 tripartite multi-entropy with the analytical prediction across representative parameter sets, Hamiltonian parameters, and inverse temperatures; we contrast the tripartite and bipartite R\'enyi-2 entropy; and we examine the structure and the fluctuations of $\rho_{AA^*}^{(2)}$ and $\rho_{AA^*}$.
We summarize and discuss our results in Section~\ref{summary_and_discussion}.

\section{Review of Tripartite Entanglement Probes}
\label{sec:quantities_of_interest}

It is only in recent years that computable probes of multipartite entnaglement in many-body systems have been constructed. Two of these probes are the \emph{reflected entropy} and \emph{multi-entropy}, which will be the primary tools with which we will characterize the tripartite entanglement structure of chaotic eigenstates. We provide a brief introduction to these quantities in this section.

\subsection{Reflected entropy}
The reflected entropy was first introduced in the context of holography \cite{Dutta2021}. 
In a tripartite Hilbert space $\mathcal{H} = \mathcal{H}_A \otimes \mathcal{H}_B \otimes \mathcal{H}_C$, the reflected entropy of a pure state $|\Psi\rangle$ is defined in terms of the canonical purification $|\rho_{AB}^{1/2} \rangle$ of the reduced density matrix $\rho_{AB} = \text{tr}_C(|\Psi\rangle\langle \Psi|)$. 
In the computational basis, we write $|\rho_{AB}^{1/2} \rangle = \sum_{a, a'}\sum_{b, b'} \langle ab|\rho_{AB}^{1/2}|a'b'\rangle |ab\rangle \otimes|a'b'\rangle$, where
$a$, $a'$, $b$, and $b'$ index complete bases of states for
$\mathcal{H}_A, \mathcal{H}_{A^*}, \mathcal{H}_B$ and $\mathcal{H}_{B^*}$, respectively. 
The reflected entropy is then defined as,
\begin{equation}
    S^R(A\!:\! B) = 
    S_{\mathrm{vN}}(\rho_{AA^*})
    =
    -\text{tr}\, \big(\rho_{AA^*} \ln \rho_{AA^*}\big),
    \label{reflected_entropy_definition}
\end{equation}
where $\rho_{AA^*}$ is the reduced density matrix of the canonical purification restricted to subsystem $\mathcal{H}_A \otimes \mathcal{H}_{A^*}$:
\begin{equation}
    \rho_{AA^*} = \text{tr}_{\mathcal{H}_B \otimes \mathcal{H}_{B^*}}\big(|\rho_{AB}^{1/2} \rangle \langle \rho_{AB}^{1/2}|\big),
    \label{rho_AA_definition}
\end{equation}
and 
$S_{\rm vN}(\rho)\equiv-\mathrm{tr}(\rho\ln\rho)$ is the von Neumann entropy associated with the density operator $\rho$.

Similar to the bipartite entanglement entropy, the reflected entropy admits R\'enyi generalizations. 
Specifically, the $(m, n)$-R\'enyi reflected entropy is defined as, 
\begin{equation}
    S^R_{m, n}(A\!:\! B) = \frac{1}{1-n} \ln \text{tr}\,\big((\rho_{AA^*}^{(m)})^n\big),
    \label{renyi_reflected_entropy_definition}
\end{equation}
where $\rho_{AA^*}^{(m)}$ is defined in direct analogy with $\rho_{AA^*}$ in Eq.\ \eqref{rho_AA_definition}, with $\rho_{AB}$ replaced by its normalized power
\begin{equation}
    \rho_{AB}^{(m)} = \frac{\rho_{AB}^m}{\text{tr}(\rho_{AB}^m)}.
    \label{rho_AB_m_definition}
\end{equation}
The canonical purification $|\rho_{AB}^{m/2}\rangle$ of $\rho_{AB}^{(m)}$ reads, in the computational basis,
\begin{equation}
    |\rho_{AB}^{m/2}\rangle = \frac{1}{\sqrt{\text{tr}(\rho_{AB}^m)}}\sum_{a, a'}\sum_{b, b'} \langle ab|\rho_{AB}^{m/2}|a'b'\rangle\, |ab\rangle \otimes|a'b'\rangle,
    \label{rho_AB_m_purification}
\end{equation}
and $\rho_{AA^*}^{(m)}$ is obtained by tracing out $\mathcal{H}_B \otimes \mathcal{H}_{B^*}$,
\begin{equation}
    \rho_{AA^*}^{(m)} = \text{tr}_{\mathcal{H}_B \otimes \mathcal{H}_{B^*}}\big(|\rho_{AB}^{m/2} \rangle \langle \rho_{AB}^{m/2}|\big).
    \label{rho_AA_m_definition}
\end{equation}
Note that we take $n \in \mathbb{Z}^+$ while $m \in 2\mathbb{Z}^+$, and we recover the von Neumann reflected entropy by analytically continuing $m, n \to 1$.
We note that apart from the $(2, n)$-R\'enyi reflected entropy, $\rho_{AA^*}^{(2)}$ also appears in the computation of 
the \emph{operator entanglement} of $\rho_{AB}$ \cite{Zanardi_2000, Zanardi_2001, Prosen_2007, Rath_2023}. 

For computational purposes, it will prove convenient to express the $(m, n)$-R\'enyi reflected entropies as expectation values of permutation operators in a replicated space.
For a pure state $|\Psi\rangle$ defined in a tripartite Hilbert space, we may write,
\begin{equation}
    S_{m, n}^R(A\!:\! B) = \frac{1}{1-n} \ln 
    \left\{
     \frac{Z^R_{m, n}(A\!:\! B)}{Z_m(AB)^n}  
     \right\}, \label{Renyi_reflected_entropy_definition_permutation}
\end{equation}
where $Z^R_{m, n}(A\!:\!B)$ and $Z_m(AB)$ are defined as 
\begin{align}
    Z_{m, n}^R(A\!:\!B) &= 
    \langle \Psi|^{\otimes mn} \, \sigma_A^{(m,n)} \otimes \sigma_B^{(m,n)}\,
    |\Psi\rangle^{\otimes mn}, 
    \nonumber \\ 
    Z_m(AB) &= \langle \Psi|^{\otimes m} \,\tau_{AB}^{(m)}\, |\Psi\rangle^{\otimes m}.
    \label{partition_function_definition}
\end{align}
Here, the permutation operators $\sigma_A$ and $\sigma_B$ both consist of $n$ $m$-cycles and act on subsystems $A$ and $B$ respectively, while $\tau_{AB}$ is an $m$-cycle acting on subsystem $AB$.
Explicitly,
\begin{align}
    \sigma_A^{(m, n)} &=  (\frac{m}{2}+1,\ldots,\frac{3m}{2})(\frac{3m}{2}+1,\ldots,\frac{5m}{2})\cdots \nonumber
    \\ & \quad \cdots (nm-\frac{m}{2}+1,\ldots,\frac{m}{2}), \nonumber \\
    \sigma_B^{(m, n)} &=  (1,\ldots, m)(m+1,\ldots, 2m)
   \nonumber \\
   & \quad  
    \cdots(nm-m+1,\ldots, nm), \nonumber \\
    \tau_{AB}^{(m)} &=  (1,2,\ldots,m),
\end{align}
with $m$ being an even number.
We note that the logarithm of the denominator is equal to $n(1-m)S_m(AB)$, where $S_m(AB) = \frac{1}{1-m} \text{tr}(\rho_{AB}^m)$ is the $m$-R\'enyi entropy of $|\Psi\rangle$ bipartitioned into $\mathcal{H}_{AB} \otimes \mathcal{H}_C$.

The reflected entropy is a promising candidate for probing tripartite entanglement. 
Indeed, the Markov gap \cite{Hayden_2021}, defined as the difference between reflected entropy and mutual information, 
$ h(A\!:\! B) = S^R(A\!:\! B) - I(A\!:\! B) $
has been proven to vanish if and only if the state being probed is a ``sum of triangle states", which can be thought of containing only bipartite 
and GHZ-type tripartite correlations \cite{Zou:2020bly, Siva_2022}. A nonzero Markov gap therefore signals the presence of tripartite correlations beyond this class.

\subsection{Multi-entropy}

Another proposed probe of tripartite entanglement is the tripartite multi-entropy
\cite{Gadde:2022cqi, Penington2023, Gadde2023, Harper:2024ker}.\footnote{In this work, the discussions on multi-entropy focus on tripartite multi-entropy. The $q$-partite multi-entropy with $q > 3$ is beyond the scope of this work.} 
The multi-entropy is a generalization of the (R\'enyi) bipartite entanglement entropy and is defined using a replica representation. 
Specifically, for a pure quantum state $|\Psi\rangle$ in a tripartite Hilbert space $\mathcal{H}_A \otimes \mathcal{H}_B \otimes \mathcal{H}_C$, the R\'enyi-$n$ tripartite multi-entropy $S_n^{(3)}$ takes the form
\begin{equation}
    S_n^{(3)} = \frac{1}{1-n}\frac{1}{n}\ln Z^{(3)}_n,
    \label{multi_entropy_definition}
\end{equation}
where $Z_n^{(3)}$ is defined as 
\begin{equation}
    Z_n^{(3)} = \langle\Psi|^{\otimes n^2} \,g_A^{(n)}\otimes g_B^{(n)}\,|\Psi\rangle^{\otimes n^2}.
\end{equation}
Here, $g_A^{(n)}$ and $g_B^{(n)}$ are both permutation operators with $n$ cycles:
\begin{align}
    \nonumber g_A^{(n)} &=  (1, 2, \ldots, n)(n+1, n+2, \ldots, n)\cdots \\ \nonumber &
    \quad \cdots (n^2-n+1, n^2-n+1, \ldots, n^2), \\ \nonumber
    g_B^{(n)} &=  (1, n+1, \ldots, n^2-n+1)(2, n+2, \ldots, n^2-n+2)\cdots \\ 
    & \quad \cdots (n, 2n,\ldots, n^2).
\end{align}
This replica definition ensures that all three subsystems are treated equally in $S_n^{(3)}$, making it a promising candidate for probing tripartite entanglement.\footnote{Due to this symmetry, we do not specify subsystems when discussing $S_n^{(3)}$ for pure states. We need to specify subsystems for $S^R_{m, n}$ since generally $S^R_{m, n}(A\!:\!B) \neq S_{m, n}^R(A\!:\!C)$.} Specifically, in random tensor networks and holographic states, it has been shown that $S_2^{(3)}$ is determined by minimal tripartitions of the system \cite{Penington2023}.

The quantity $S_n^{(3)}$ is sensitive to both tripartite \emph{and} bipartite entanglement. 
To separate out tripartite entanglement, a quantity called genuine multi-entropy has been proposed \cite{Penington2023, Iizuka:2025ioc, Iizuka:2025caq}, which is defined as, 
\begin{equation}
    GM_n^{(3)} = S_n^{(3)} - \frac{1}{2}(S_n(A)+S_n(B) + S_n(C)).
    \label{genuine_multi_entropy_definition}
\end{equation}
Here, $S_n(A)$ is the bipartite R\'enyi-$n$ entropy. The genuine multi-entropy has been studied in black holes and holographic settings \cite{Penington2023, Iizuka:2025ioc, Iizuka:2025caq}. 

We note that the R\'enyi-$2$ tripartite multi-entropy is related to the R\'enyi-$2$ tripartite multi-entropy \cite{Liu_2025}:
\begin{equation}
    S_2^{(3)} = \frac{1}{2}S_{2, 2}^R(A\!:\!B) + S_2(AB).
    \label{multi_reflected_relation}
\end{equation}
Moreover, we may define a R\'enyi generalization of the Markov gap $h_{m, n}(A\!:\!B) = S_{m, n}^R(A\! :\! B) - I_n(A\!:\! B)$,
where $I_n(A\!:\!B) = S_n(A) + S_n(B) - S_n(AB)$.
For $m=n=2$,
this quantity reduces to the genuine multi-entropy:
$GM_2^{(3)} = (1/2) h_{2,2}(A\!:\! B)$.

\section{Tripartite Entanglement Structure of Constrained Haar-Random States}
\label{section_2}

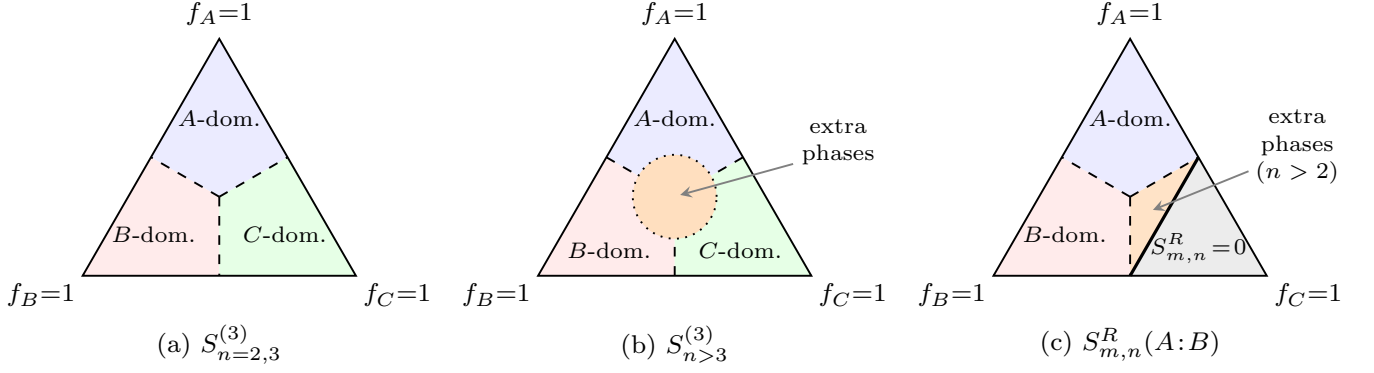
\begin{figure*}[t]
    \centering
    \resizebox{\textwidth}{!}{%
    \begin{tikzpicture}[>=stealth, line width=0.6pt, font=\footnotesize]
    \def\Ax{0}\def\Ay{2.598}\def\Bx{-1.5}\def\By{0}\def\Cx{1.5}\def\Cy{0}
    \def\panelsep{5.0}

    \begin{scope}[shift={(0,0)}]
      \coordinate (A) at (\Ax,\Ay); \coordinate (B) at (\Bx,\By); \coordinate (C) at (\Cx,\Cy);
      \coordinate (Gc) at (0,0.866);
      \coordinate (mAB) at ($ (A)!0.5!(B) $); \coordinate (mBC) at ($ (B)!0.5!(C) $); \coordinate (mAC) at ($ (A)!0.5!(C) $);
      \fill[blue!8] (A)--(mAB)--(Gc)--(mAC)--cycle;
      \fill[red!8] (B)--(mBC)--(Gc)--(mAB)--cycle;
      \fill[green!10] (C)--(mAC)--(Gc)--(mBC)--cycle;
      \draw (A)--(B)--(C)--cycle;
      \draw[dashed] (Gc)--(mAB); \draw[dashed] (Gc)--(mBC); \draw[dashed] (Gc)--(mAC);
      \node[above=1pt] at (A) {$f_A{=}1$}; \node[below left=-1pt] at (B) {$f_B{=}1$}; \node[below right=-1pt] at (C) {$f_C{=}1$};
      \node at ($ (A)!0.5!(Gc) $) {\scriptsize $A$-dom.};
      \node at ($ (B)!0.52!(Gc) $) {\scriptsize $B$-dom.};
      \node at ($ (C)!0.52!(Gc) $) {\scriptsize $C$-dom.};
      \node at (0,-0.75) {(a) $S_{n=2, 3}^{(3)}$};
    \end{scope}

    \begin{scope}[shift={(\panelsep,0)}]
      \coordinate (A) at (\Ax,\Ay); \coordinate (B) at (\Bx,\By); \coordinate (C) at (\Cx,\Cy);
      \coordinate (Gc) at (0,0.866);
      \coordinate (mAB) at ($ (A)!0.5!(B) $); \coordinate (mBC) at ($ (B)!0.5!(C) $); \coordinate (mAC) at ($ (A)!0.5!(C) $);
      \fill[blue!8] (A)--(mAB)--(Gc)--(mAC)--cycle;
      \fill[red!8] (B)--(mBC)--(Gc)--(mAB)--cycle;
      \fill[green!10] (C)--(mAC)--(Gc)--(mBC)--cycle;
      \fill[orange!25] (Gc) circle (0.46); \draw[dotted] (Gc) circle (0.46);
      \coordinate (bAB) at ($(Gc)!0.531!(mAB)$);
      \coordinate (bBC) at ($(Gc)!0.531!(mBC)$);
      \coordinate (bAC) at ($(Gc)!0.531!(mAC)$);
      \draw (A)--(B)--(C)--cycle;
      \draw[dashed] (mAB)--(bAB); \draw[dashed] (mBC)--(bBC); \draw[dashed] (mAC)--(bAC);
      \node[above=1pt] at (A) {$f_A{=}1$}; \node[below left=-1pt] at (B) {$f_B{=}1$}; \node[below right=-1pt] at (C) {$f_C{=}1$};
      \node at ($ (A)!0.5!(Gc) $) {\scriptsize $A$-dom.};
      \node at ($ (B)!0.52!(Gc)+(0,-0.17) $) {\scriptsize $B$-dom.};
      \node at ($ (C)!0.52!(Gc)+(0,-0.17) $) {\scriptsize $C$-dom.};
      \node[align=center,fill=white,inner sep=1pt,anchor=west] at ($(Gc)+(1.35,0.60)$) {\scriptsize extra\\[-1pt]\scriptsize phases};
      \draw[->,gray] ($(Gc)+(1.30,0.36)$) -- ($(Gc)+(0.06,0.02)$);
      \node at (0,-0.75) {(b) $S_{n>3}^{(3)}$};
    \end{scope}

    \begin{scope}[shift={(2*\panelsep,0)}]
      \coordinate (A) at (\Ax,\Ay); \coordinate (B) at (\Bx,\By); \coordinate (C) at (\Cx,\Cy);
      \coordinate (mAC) at ($ (A)!0.5!(C) $); \coordinate (mBC) at ($ (B)!0.5!(C) $); \coordinate (mAB) at ($ (A)!0.5!(B) $);
      \coordinate (Gc) at (0,0.866);
      \coordinate (mExtra) at ($(mAC)!0.5!(mBC)$);
      \coordinate (ExtraCenter) at ($(Gc)!0.6667!(mExtra)$);
      \coordinate (mGray) at ($(mAC)!0.5!(mBC)$);
      \coordinate (GrayCenter) at ($(C)!0.6667!(mGray)$);
      \fill[gray!15] (C)--(mAC)--(mBC)--cycle;
      \fill[blue!8] (A)--(mAC)--(Gc)--(mAB)--cycle;
      \fill[red!8] (B)--(mBC)--(Gc)--(mAB)--cycle;
      \fill[orange!22] (mAC)--(mBC)--(Gc)--cycle;
      \draw (A)--(B)--(C)--cycle;
      \draw[line width=1.0pt] (mAC)--(mBC);
      \draw[dashed] (Gc)--(mAB); \draw[dashed] (Gc)--(mAC); \draw[dashed] (Gc)--(mBC);
      \node[above=1pt] at (A) {$f_A{=}1$}; \node[below left=-1pt] at (B) {$f_B{=}1$}; \node[below right=-1pt] at (C) {$f_C{=}1$};
      \node at ($ (A)!0.5!(Gc) $) {\scriptsize $A$-dom.}; \node at ($ (B)!0.52!(Gc) $) {\scriptsize $B$-dom.};
      \node[inner sep=0pt] at ($(GrayCenter)+(0,-0.13)$) {\scriptsize $S^R_{m,n}\!=\!0$};
      \node[align=center,fill=white,inner sep=0.5pt,anchor=west] at ($(Gc)+(1.35,0.55)$) {\scriptsize extra\\[-1pt]\scriptsize phases\\[-1pt]\scriptsize $(n>2)$};
      \draw[->,gray] ($(Gc)+(1.30,0.28)$) -- (ExtraCenter);
      \node at (0,-0.75) {(c) $S^R_{m,n}(A\!:\!B)$};
    \end{scope}
    \end{tikzpicture}}
    \caption{Entanglement phase diagrams of the (a) $n=2$ and (b) $n>2$ R\'enyi multi-entropies as well as the (c) R\'enyi reflected entropy in a constrained Haar-random state, drawn on the simplex of subsystem fractions $f_K=V_K/V$ ($f_A+f_B+f_C=1$). The phases are labeled by which subregion has the dominant contribution to the corresponding entanglement quantity. There are exactly three phases in (a), separated by the medians $f_A=f_B$, $f_B=f_C$, $f_A=f_C$. The same phases appear in (b), but additional phases can arise near the center where all three subsystems are comparable, which we do not characterize. In (c), the thick line $V_A+V_B=V_C$ (i.e.\ $f_C=1/2$) bounds the region where the leading-order reflected entropy vanishes. }
    \label{fig:phase-diagram}
\end{figure*}

We now turn to a detailed analysis of the constrained Haar-random ansatz of Eq.~\eqref{ergodic_tripartition} through the lens of the tripartite entanglement probes reviewed above. In particular, we obtain analytical leading order expressions for the average R\'enyi multi-entropy and R\'enyi reflected entropy and, assuming a particular density of states, obtain explicit expressions using a saddle-point approximation. As functions of subsystem sizes, these quantities exhibit the entanglement phase diagrams shown in Fig.~\ref{fig:phase-diagram}. While the topologies of the phase diagrams are the same as for unconstrained Haar states 
\cite{Penington2023, Akers2022, Iizuka:2024pzm},
the subleading corrections to the volume law scaling are different. As we will explain, the analytic structure of the expressions for the R\'enyi multi-entropies and reflected entropies differ qualitatively from that of the unconstrained case.

As a more fine-grained probe of the tripartite entanglement structure of these states, we analytically compute the structure of the full reduced density matrix of the canonical purification, $\rho_{AA^*}^{(2)}$. Intriguingly, by tuning relative subsystem sizes, $\rho_{AA^*}^{(2)}$ interpolates between three regimes closely related to the references states described in Refs.~\cite{Verlinde2020,Verlinde2021} in the context of a two-sided eternal black hole: the product state, the thermo-\emph{mixed} double, and the thermofield double. Our claim is that the universal structure of $\rho_{AA^*}^{(2)}$ for general chaotic eigenstates is captured by  constrained Haar-random states. This may be understood as a tripartite generalization of subsystem ETH \cite{Dymarsky_2018}, which proposes a similar relation for the bipartite reduced density matrix. We will expand on this connection and provide evidence in favor of this claim via numerical studies in Sec.\ \ref{sec:exact_diagonalization_multi_entropy}.

\subsection{R\'enyi multi-entropy}
\label{Renyi_multi_entropy_analytical}

We begin by studying the tripartite R\'enyi multi-entropy; as we will see, the strategies we employ in this computation will also prepare us for the computation of the R\'enyi reflected entropies in the subsequent subsection.
Specifically, in this section, we present the analytical prediction for 
$\average{Z^{(3)}_n}$, 
the ensemble average of $Z_n^{(3)}$ for the ansatz of Eq.~\eqref{ergodic_tripartition}. 
Here, we use $\average{S}$ to denote the ensemble average of a quantity $S$.
In our calculation, we take the window function $F(E_i^A+E_j^B+E_k^C-E)$ to be a delta function peaked at $E_i^A + E_j^B + E_k^C = E$. 
We then approximate $\average{S_n^{(3)}}$ as $\frac{1}{1-n}\frac{1}{n} \ln (\average{Z_n^{(3)}})$. 

First, we outline the method of calculation, using $\average{Z_2^{(3)}}$ as an example. 
Starting from the replica-trick definition of multi-entropy in Eq.~\eqref{multi_entropy_definition}, $\average{Z_n^{(3)}}$ is equivalent to the average of the product of $n^2$ coefficients $M_{ijk}$ and their complex conjugates $M_{ijk}^*$, with indices contracted according to the permutation operators $g_A^{(n)}$, $g_B^{(n)}$, and an implicit identity permutation operator for subsystem $C$. 
When $n = 2$,
\begin{widetext}
\begin{align}
    \average{Z_2^{(3)}} & =  \average{\sum_{\substack{i_1,\ldots, i_4;i_1',\ldots, i_4';\\ j_1,\ldots, j_4;j_1',\ldots j_4';\\k_1,\ldots, k_4}} M_{i_1j_1k_1}M^*_{i_1'j_1'k_1}M_{i_2j_2k_2}M_{i_2'j_2'k_2}M_{i_3j_3k_3}M^*_{i_3'j_3'k_3}M_{i_4j_4k_4}M_{i_4'j_4'k_4}\delta_{i_1i_2'}\delta_{i_2i_1'}\delta_{i_3i_4'}\delta_{i_4i_3'}\delta_{j_1j_3'}\delta_{j_3j_1'}\delta_{j_2j_4'}\delta_{j_4j_2'}} 
    \label{Z_2_3_product_Mijk}
    \\ \nonumber
    & = \sum \average{M_{i_1j_1k_1} M^*_{i_2j_3k_1}M_{i_2j_2k_2}M^*_{i_1j_4k_2}M_{i_3j_3k_3}M^*_{i_4j_1k_3}M_{i_4j_4k_4}M^*_{i_3j_2k_4}}.
\end{align}
\end{widetext}
Here,
$\delta_{i_1i_2'}\delta_{i_2i_1'}\delta_{i_3i_4'}\delta_{i_4i_3'}$ comes from the permutation operator $g_A^{(2)}$, and $\delta_{j_1j_3'}\delta_{j_3j_1'}\delta_{j_2j_4'}\delta_{j_4j_2'}$ comes from $g_B^{(2)}$. 
The $k$ indices are the same for each pair of $M_{ijk}$ and $M^*_{i'j'k}$ due to the implicit identity permutation for $\mathcal{H}_C$.

The average of this product can be expressed as a sum of Wick contractions, and each Wick contraction is associated with a distinct permutation operator representing how $M_{ijk}$ and $M_{i'j'k'}^*$ are paired together in the contraction. 
Thus, we can write,
\begin{equation}
    \average{Z_n^{(3)}} = \sum_\tau Z_n^{(3)}(\tau),
\end{equation}
where $Z_n^{(3)}(\tau)$ denotes the contribution from the Wick contraction associated with the permutation operator $\tau$:
\begin{equation}
    Z_n^{(3)}(\tau) = \sum \prod_{\mu=1}^{n^2} \average{M_{i_\mu j_\mu k_\mu}M_{i_{(g_A^{(n)}\circ\tau^{-1})(\mu)}j_{(g_B^{(n)}\circ\tau^{-1})(\mu)}k_{(\tau^{-1})(\mu)}}^*}.
    \label{permutation_operator_expression}
\end{equation}

Unlike the case of Haar random states, 
the ansatz
\eqref{ergodic_tripartition} is a sum of product states of subsystem energy eigenstates $|E_i^A\rangle|E_j^B\rangle|E_k^C\rangle$ constrained by energy conservation. 
As a result, the Wick contractions produce independent sums over the conserved charges (i.e., subsystem energies). 
Specifically, each Wick contraction organizes into independent replica loops, and each loop is associated with a different energy $E^K_i$ (the subsystem is labeled by $K$). A replica loop associated with $E^K_i$ contributes a factor of $e^{S_K(E^K_i)}$, which is the density of states at the energy $E^K_i$. 
The sums over two of the subsystem energies are independent, while the third is fixed by energy conservation $E^A + E^B + E^C = E$. 

In the thermodynamic limit, the leading order contribution to $\average{Z_n^{(3)}}$ is determined by the permutations, $\tau$, that maximize $Z_n^{(3)}(\tau)$ [Eq.\ \eqref{permutation_operator_expression}]. 
Assuming that the entropy density is the same for each subsystem, the microcanonical entropy $S_K(E^K)$ scales with system size $S_K(E^K) \propto V_K$ ($K = A, B, C$). 
Thus, when one system is much larger than the others (e.g., $V_C \gg V_A, V_B$), the dominant permutation operator is the one that maximizes the number of replica loops (and thus, the number of $S_K(E^K)$ in the exponent in $\average{Z_n^{(3)}}$) corresponding to that subsystem (e.g., the identity permutation that gives $n^2$ replica loops corresponding to subsystem $C$). 
When two or more subsystems are of comparable sizes, on the other hand, there can be dominant permutation operators that do not maximize any of the three individual cycle numbers but maximize $Z_n^{(3)}(\tau)$, as we will see later in the computation of R\'enyi reflected entropy. 
The dominant permutation operators in the calculation of R\'enyi multi-entropy have been discussed in the case of Haar random states \cite{Iizuka:2024pzm}.

The leading-order contribution to $\average{Z_n^{(3)}}$ for a chaotic eigenstate $|E\rangle$ with energy $E$ takes the following forms depending on the relative sizes of the subsystems:
\begin{widetext}
\begin{equation}
    \average{Z_n^{(3)}} = 
     \frac{1}{\mathcal{N}^{n^2}} \begin{cases}
       \sum_{E^A_1,\ldots, E^A_n}\sum_{E^B_1,\ldots, E^B_n} e^{\sum_{i=1}^nS_A(E^A_i) + \sum_{j=1}^nS_B(E^B_j) + \sum_{i,j=1}^nS_C(E - E^A_i - E^B_j)}, \quad V_C \gg V_A, V_B \\
        \sum_{E^A_1,\ldots, E^A_n}\sum_{E^C_1,\ldots, E^C_n} e^{\sum_{i=1}^nS_A(E^A_i) + \sum_{i,j=1}^nS_B(E - E^A_i - E^C_j) + \sum_{j=1}^nS_C(E^C_j)},
        \quad V_B \gg V_A, V_C \\ 
        \sum_{E^B_1,\ldots, E^B_n}\sum_{E^C_1,\ldots, E^C_n} e^{\sum_{i,j=1}^nS_A(E - E^B_i - E^C_j) + \sum_{i=1}^nS_B(E^B_i) + \sum_{j=1}^nS_C(E^C_j)}, 
        \quad V_A \gg V_B, V_C
    \end{cases}
    \label{Z_n_3 average}
\end{equation}
The corresponding dominant permutation operators are $e$ (the identity), $g_B^{(n)}$ and $g_A^{(n)}$ respectively.
Here, 
\begin{equation}
    \mathcal{N} = \sum_{E^A} \sum_{E^B} e^{S_A(E^A) + S_B(E^B) + S_C(E - E^A - E^B)}
    \label{normalization constant}
\end{equation}
is the normalization coefficient.

Performing the saddle point approximation for Eq.\ \eqref{Z_n_3 average}, we find that $\average{Z_n^{(3)}}$ is dominated by the \emph{replica-symmetric} 
contribution; we show this explicitly in Appendix \ref{appendix:symmetric_saddle_multi_entropy}).
By replica symmetry we mean that the dominant saddle-point energies are independent of the replica index, i.e., $E_i^A = E^A$ and $E_j^B = E^B$ for all $i,j$, so that the $n^2$ replicas reduce to a single effective energy for each subsystem.
Therefore, we present a replica-symmetric approximation to $\average{Z_n^{(3)}}$, which we denote as $\average{Z_{n}^{(3)}}_{\text{RS}}$ (``RS" denotes replica-symmetric):
\begin{align}
    \average{Z_{n}^{(3)}}_{\text{RS}} = 
    \frac{1}{\mathcal{N}^{n^2}} \begin{cases}
        \sum_{E^A}\sum_{E^B} e^{nS_A(E^A) + nS_B(E^B) + n^2S_C(E - E^A - E^B)}, \quad V_C \gg V_A, V_B \\
        \sum_{E^A}\sum_{E^B} e^{nS_A(E^A) + n^2S_B(E^B) + nS_C(E-E^A-E^B)}, 
        \quad V_B \gg V_A, V_C \\ 
        \sum_{E^A}\sum_{E^B} e^{n^2S_A(E^A) + nS_B(E^B) + nS_C(E - E^A - E^B)}, \quad V_A \gg V_B, V_C
    \end{cases}
    \label{Z_n_3_average_n_approx}
\end{align}
\end{widetext}
In this form, the relative weights each subsystem contributes to the entropy under the permutation $\tau$ are immediately visible from the exponents. 
Moreover, $\average{Z_{n}^{(3)}}_{\text{RS}}$ corresponds cleanly to the expressions for R\'enyi multi-entropy in Haar random states \cite{Iizuka:2024pzm}. 
For example, in the regime where $V_C \gg V_A, V_B$, $Z_n^{(3)}$ for Haar random states take the form $2^{nV_A + nV_B + n^2 V_C}/\mathcal{N}_{\text{Haar}}^{n^2}$, where $\mathcal{N} = 2^{V_A + V_B + V_C}$ is the appropriate normalization.

Since $\average{Z_{n}^{(3)}}_{\text{RS}}$ shares the same saddle point as $\average{Z_n^{(3)}}$, $\average{Z_{n}^{(3)}}_{\text{RS}}$ correctly captures the leading-order, $O(V)$, behavior of R\'enyi multi-entropy in the thermodynamic limit. 
At subleading order, on the other hand, $\average{Z_{n}^{(3)}}$ exhibits a $O(\ln V)$ negative departure from $\average{Z_{n}^{(3)}}_{\text{RS}}$. 
We present a derivation of this $O(\ln V)$ difference assuming Gaussian density of states in Appendix \ref{appendix:multi_entropy_exact_RS_comparison}. 
A Gaussian density of states is a reasonable approximation for systems with symmetric microcanonical entropy density $s(u) = s(-u)$ close to infinite temperature. In particular, it approximates the mid-spectrum density of states of the mixed field Ising model (which we study in Sec.~\ref{sec:exact_diagonalization_multi_entropy}) well in the thermodynamic limit.
As shown in Appendix \ref{appendix:multi_entropy_exact_RS_comparison}, the $O(\ln V)$ difference stems from the fluctuations in the replica subsystem energies $E_i^K$ around their respective averages $\average{E^K}$ in $\average{Z_{n}^{(3)}}$. 

We note that this $O(\ln V)$ difference between $\average{Z_{n}^{(3)}}$ and $\average{Z_{n}^{(3)}}_{\text{RS}}$ does not indicate a $O(\ln V)$ negative departure of $\average{Z_{n}^{(3)}}$ from the Haar random value. 
In fact, as shown in Appendix \ref{appendix:multi_Haar_random_comparison}, under the approximation of Gaussian density of states, for states in the middle of the spectrum (with inverse temperature $\beta = 0$), this $O(\ln V)$ difference exactly cancels the $O(\ln V)$ positive departure of $\average{Z_{n}^{(3)}}_{\text{RS}}$ from the Haar random value.
In other words, taking into account the fluctuations in the replica subsystem energies corrects the spurious $O(\ln V)$ positive departure from the random Haar value exhibited by the replica-symmetric expression.
Similar to the R\'enyi bipartite entropies with $n > 1$ and the von Neumann entropy  of mid-spectrum eigenstates \cite{Lu_2017, Rodriguez_Nieva_2024, Haque_2022, Huang_2019, Huang_2021, Huang_2024}, $\average{Z_{n}^{(3)}}$ with $n>1$ exhibits an $O(1)$ departure from the random Haar value.\footnote{When $E$ scales with $V$, on the other hand, we would generally expect the entanglement measures to deviate more from the Haar random value.}
Concretely, at mid-spectrum the fluctuation-corrected multi-entropy departs from the Haar value $\average{S^{(3)}_{n,\text{Haar}}}=V_A+V_B$ only through a partition-dependent $O(1)$ term: writing $f_K=V_K/V$, the $\beta=0$ limit of the full expression derived in Appendix~\ref{appendix:multi_Haar_random_comparison} reads
\begin{widetext}
\begin{align}
  \average{S_n^{(3)}}\big|_{\beta=0} = V_A+V_B
   + 
   \frac{1}{n(1-n)}\ln\!\Bigg\{\frac{f_C^{\,n-1}}{f_C^{(n^2-1)/2}(nf_A+nf_B+f_C)^{1/2}\big[(nf_A+f_C)(nf_B+f_C)\big]^{(n-1)/2}}\Bigg\},
  \label{eq:O1_multi_maintext}
\end{align}
\end{widetext}
whose logarithmic term is $O(1)$, is not constant across partitions, and is negative (so that $\average{S_n^{(3)}}<\average{S^{(3)}_{n,\text{Haar}}}$). This is Eq.~\eqref{appendix_multi_entropy_O1}; an analogous $O(1)$ term holds for the reflected entropy, as we discuss in the following subsection.

The difference between \(\average{Z_n^{(3)}}\) and \(\average{Z_n^{(3)}}_{\rm RS}\) can be interpreted as a departure from the simple cycle-counting structure of the unconstrained Haar-random calculation. Computationally, $\average{Z_{n}^{(3)}}_{\text{RS}}$ takes a ``cycle-counting" form, where the exponent of the subsystem Hilbert space dimension (or density of states in the case with energy conservation) is equal to the number of cycles in the composition of the subsystem permutation operator (i.e., $g_A^{(n)}$, $g_B^{(n)}$, and the identity operator for subsystem $C$) and the permutation operator associated with the dominant Wick contraction [see Eq.\ \eqref{permutation_operator_expression}]. The ``cycle-counting" form is an exact expression for the corresponding Wick contraction if all Wick contraction loops are associated with the same subsystem volume, which is true in the case of random Haar states with no constraint, or when computing the leading order expression for R\'enyi bipartite entropy where there is only one replica loop for the smaller subsystem (when $V_A < V_B$, the leading order term for R\'enyi bipartite entropy is $\sum_{E^A} e^{S_A(E^A) + nS_B(E - E^A)}$).
When computing multipartite entanglement quantities in random Haar states with symmetry constraints, the fact that each replica loop is associated with a distinct symmetry charge affects the subleading contribution.

Finally, we note that when $n=2$ and $n=3$, there are only three possible dominating permutations (depending on which subsystem is the largest), and the ``much greater than" sign in Eq.\ \eqref{Z_n_3_average_n_approx} becomes a simple ``greater than" sign:
\begin{widetext}
\begin{equation}
    \average{Z_n^{(3)}} = 
    \frac{1}{\mathcal{N}^{n^2}} \begin{cases}
        \sum_{E^A}\sum_{E^B} e^{nS_A(E^A) + nS_B(E^B) + n^2S_C(E - E^A - E^B)}, 
        \quad V_C > V_A, V_B \\
        \sum_{E^A}\sum_{E^B} e^{nS_A(E^A) + n^2S_B(E^B) + nS_C(E-E^A-E^B)}, \quad V_B > V_A, V_C \\ 
        \sum_{E^A}\sum_{E^B} e^{n^2S_A(E^A) + nS_B(E^B) + nS_C(E - E^A - E^B)}, \quad V_A > V_B, V_C
    \end{cases}
    \label{Z_n_3 average_n_2_3}
\end{equation}
\end{widetext}
When $n \geq 4$, however, there can be additional phases in the region $V_A \sim V_B \sim V_C$ with dominating contributions that take different forms from the three cases in the above expression. 
For example, when $n=4$, $\frac{1}{\mathcal{N}^{16}}\sum_{E^A}\sum_{E^B}e^{7S_A(E^A)+7S_B(E^B)+11S_C(E-E^A-E^B)}$ is one of the dominating terms that arise in the region $V_A \sim V_B \sim V_C$. 
The exact form of the leading order contribution to $\average{Z_n^{(3)}}$ in the above-mentioned region for general $n \geq 4$ remains an open question for both Haar random states without and with constraints. 

\subsubsection{Curvature of R\'enyi multi-entropy}

We briefly discuss the curvature of the multi-entropy in this subsection.
Recall that for Haar random states in a bipartite Hilbert space $\mathcal{H}_A \otimes \mathcal{H}_B$, at leading order, the R\'enyi-$n$ entropy takes the form $\average{S_n(A)} = V_A$ when $V_A < V_B$, and thus depends linearly on $f = V_A/V$. 
However, it has been shown \cite{Lu_2017} that in chaotic eigenstates, when the R\'enyi index $n > 1$, the leading order average R\'enyi entropy $\average{S_n(A)}$ is a convex function of $f$, i.e., $\frac{\partial^2 \average{S_n(A)}}{\partial f^2} > 0$. This feature allows R\'enyi entropy to distinguish pure eigenstates of a chaotic system from thermal states at the same energy \cite{Garrison_2018, Lu_2017}.

In Appendix \ref{appendix:multi_entropy_curvature}, we examine the curvature of the R\'enyi multi-entropy. In the regime where $V_C \gg V_A, V_B$, $\average{S_n^{(3)}}$ can be treated as a function of two independent variables $f_A = V_A/V$ and $f_B = V_B/V$. For $n > 1$, we find that $\average{S_n^{(3)}}$ is flat (i.e., the first derivative vanishes) along the $f_A - f_B$ direction, indicating that $\average{S_n^{(3)}}$ is insensitive to the internal partition of $\mathcal{H}_A \otimes \mathcal{H}_B$. 
This behavior is consistent with the property of $\average{S_n^{(3)}}$ in Haar random states.
On the other hand, $\average{S_n^{(3)}}$ is convex along the $f_A + f_B$ direction, mirroring the convexity of R\'enyi entropy, as is visible in Fig.\ \ref{fig:multi_entropy_cusp_draft}.
This convexity is consistent with the interpretation of R\'enyi multi-entropy as a natural multipartite generalization of R\'enyi entropy.

\subsection{R\'enyi reflected entropy}
\label{Renyi_reflected_entropy_analytical}

Based on 
the ansatz
\eqref{ergodic_tripartition}, and starting from the replica-trick definition of R\'enyi reflected entropy 
\eqref{renyi_reflected_entropy_definition}, we can calculate $\average{Z_{m,n}^R}$ using the same method as outlined in Sec.~\ref{Renyi_multi_entropy_analytical}. 
Here, we directly present the calculation results. 
In the thermodynamic limit, for a chaotic eigenstate $|E\rangle$ with energy $E$, the leading-order Wick contraction gives
one of the four forms below,
depending on the subsystem sizes.
These regimes are the phases
of the R\'enyi reflected entropy 
in
Fig.\ \ref{fig:phase-diagram}(c).
The first case $V_A+B_B<V_C$
is the region where $C$ is large enough to purify $AB$,  leading to a vanishing leading-order reflected entropy; 
the second and third are
the $A$- and $B$-dominated regions;
and the fourth, 
present only for $n>2$,
occupies the central region 
where all three subsystems are of 
comparable size.
Explicitly, in these four regions, we have for the R\'enyi reflected entropy,
\begin{widetext}
\begin{enumerate}
\item
$\text{For} \ V_A + V_B  < V_C:$ 
\begin{align}
\label{Z_m_n_R_exact} 
    \average{Z_{m,n}^R} &= 
    \frac{1}{\mathcal{N}^{mn}} \sum_{E^A_1,\ldots, E^A_n} \sum_{E^B_1,\ldots, E^B_n} 
    e^{\sum_{i=1}^nS_A(E^A_i) + \sum_{i=1}^n S_B(E^B_j)}
    \nonumber \\
    &\quad \times
    e^{\frac{m}{2}\sum_{i=1}^n\{S_C(E - E^A_i - E^B_i)  
    +S_C(E - E^A_i - E^B_{(i+1) \bmod n})\}};  
\end{align}
\item
$\text{For} \ V_A + V_B  > V_C, V_A > V_B, V_A > V_C + (1 - \frac{2}{n})V_B:$
\begin{align}
\average{Z_{m,n}^R} &=   \frac{1}{\mathcal{N}^{mn}} \sum_{E^B_i,\ldots, E^B_{mn+2-2n}} \sum_{E^C_1, \ldots, E^C_n} e^{\sum_{i=1}^{mn+2-2n} S_B(E^B_i) + \sum_{j=1}^n S_C(E^C_j)}
\nonumber \\
&\quad \times e^{\sum_{j=1}^n
\{S_A(E - E^B_1 - E^C_j)  
+S_A(E - E^B_2 - E^C_j)\} 
+ \sum_{j=1}^n\sum_{i=1}^{m-2} S_A(E - E^B_{(m-2)(j-1) + i + 2} - E^C_j)}; 
\end{align}
\item
$\text{For} \ V_A + V_B  > V_C, V_B > V_A, V_B > V_C + (1 - \frac{2}{n})V_A:$
\begin{align}
\average{Z_{m,n}^R} &=   \frac{1}{\mathcal{N}^{mn}} \sum_{E^A_i,\ldots, E^A_{mn+2-2n}} \sum_{E^C_1, \ldots, E^C_n} 
e^{\sum_{i=1}^{mn+2-2n} S_A(E^A_i) + \sum_{j=1}^n S_C(E^C_j)}
\nonumber \\
&\quad \times 
e^{\sum_{j=1}^n \{S_B(E - E^A_1 - E^C_j)  
+S_B(E - E^A_2 - E^C_j)\}
+ \sum_{j=1}^n\sum_{i=1}^{m-2} S_B(E - E^A_{(m-2)(j-1) + i + 2} - E^C_j)}; 
\end{align}
\item
Otherwise:
\begin{align}
\average{Z_{m,n}^R} &=   \frac{1}{\mathcal{N}^{mn}} \sum_{E^B_1, \ldots, E^B_{m(n-1)}} \sum_{E^C_1, \ldots, E^C_{2n}} 
e^{\sum_{i=1}^{m(n-1)} S_B(E^B_i) + \sum_{j=1}^{2n} S_C(E^C_j) + \sum_{i=1}^n S_A(E - E^B_i - E^C_{2i}) }
\nonumber \\
&\quad \times 
e^{
\sum_{j=1}^{2n} \sum_{i=1}^{m-1} S_A(E - E^B_{(m-1)(j-1) + i + n} - E^C_j) 
}
\prod_{i=1}^{n} \delta(E^C_{2i} + E^B_i - E^C_{(2i+1) \bmod 2n} - E^B_{(i+1) \bmod n}).
\end{align}
\end{enumerate}
Here, $\mathcal{N}$ takes the form as expressed in Eq. \eqref{normalization constant}. 

These forms come directly from Wick contractions corresponding to dominant permutations:
\begin{equation}
    Z_{m, n}^R(\tau) = \sum \prod_{\mu=1}^{n^2} \average{M_{i_\mu j_\mu k_\mu}M_{i_{(\sigma_A^{(m,n)}\circ\tau^{-1})(\mu)}j_{(\sigma_B^{(m,n)}\circ\tau^{-1})(\mu)}k_{(\tau^{-1})(\mu)}}^*}.
\end{equation}
The dominant permutation operators corresponding to the first three cases are $e$ (the identity), $\sigma_A^{(n)}$ and $\sigma_B^{(n)}$ respectively. 
The last case is only relevant if $n > 2$, and the dominant permutation operator takes the form $X = (1, \ldots, m/2)(m/2+1,\ldots,m)\cdots(mn-m/2+1,\ldots,mn)$, which consists of $2n$ cycles of length $m/2$. 
These dominant permutation operators were determined in \cite{Akers2022}.

The denominator in Eq.\ \eqref{Renyi_reflected_entropy_definition_permutation} is simply the partition function for the $m$-R\'enyi bipartite entropy, which takes the form
\begin{equation}
    \average{Z_m} = \frac{1}{\mathcal{N}^m} \begin{cases}
    \sum_{E^{AB}} e^{S_{AB}(E^{AB}) + mS_C(E - E^{AB})}, \quad V_{AB} < V_C \\
    \sum_{E^{AB}} e^{mS_{AB}(E^{AB}) + S_C(E - E^{AB})}, \quad V_{AB} > V_C
    \end{cases}
\end{equation}
Here $e^{S_{AB}(E^{AB})}$ is the density of states in subsystem $\mathcal{H}_A \otimes \mathcal{H}_B$ at energy $E^{AB} = E^A + E^B$, and $V_{AB} = V_A + V_B$ is the size of subsystem $\mathcal{H}_A \otimes \mathcal{H}_B$. 

It is natural to wonder whether $\average{Z_{m, n}^R}$ can be written in a replica-symmetric form similar to $\average{Z_n^{(3)}}$. 
Analyzing the structure of Eq.\ \eqref{Z_m_n_R_exact}, we find that $\average{Z_{m, n}^R}$ is dominated by the replica-symmetric saddle in the regime $V_A + V_B < V_C$.
Noting that the exponent in $\average{Z_{m, n}^R}$ in this regime is symmetric under the group of cyclic transformations $C_n$ generated by $g: (E_i^A, E_i^B) \to (E_{(i+1) \text{mod} \ n}^A, E_{(i+1) \text{mod} \ n}^B)$, the replica symmetry of the saddle point can be derived following the exact same argument as that for R\'enyi multi-entropy presented in Appendix \ref{appendix:symmetric_saddle_multi_entropy}.
This allows the approximation of $\average{Z_{m, n}^R}$ with the simpler replica-symmetric form
    \begin{equation}
        \average{Z_{m, n}^R}_{\text{RS}} = \frac{1}{\mathcal{N}^{mn}}\sum_{E^A}\sum_{E^B} e^{nS_A(E^A) + nS_B(E^B) + mnS_C(E - E^A - E^B)},
        \quad V_A + V_B < V_C
    \end{equation}
\end{widetext}

However, in the other regimes, $\average{Z_{m, n}^R}$ does not share the same saddle point as the corresponding replica-symmetric expressions.
For example, in the regime where $V_A + V_B  > V_C$, $V_A > V_B$, and $V_A > V_C + (1 - \frac{2}{n})V_B$, $E_1^B$ and $E_2^B$ have different weights in the exponent compared to the other $E_i^B$, and these two types of $E^B$s take different values at the saddle point. 
This would lead to deviations of the replica-symmetric expression from $\average{Z_{m, n}^R}$ beyond order $O(\ln V)$.
In fact, assuming Gaussian density of states, apart from a $O(\ln V)$ difference, $\average{S_{m, n}^R}$ and the corresponding replica-symmetric expression in this regime exhibit an additional $O(E^2/V)$ difference. 
Though $O(E^2/V)$ would be at most $O(1)$ under the assumptions required for the density of states to be approximated as Gaussian, we expect the difference to be more relevant when the energy scales faster than $\sqrt{V}$. 

Finally, we briefly discuss the saddle point approximation to $\average{S_{m, n}^R}$ in the regime $V_A + V_B < V_C$. 
In states of the form described by Eq.\ \eqref{ergodic_tripartition}, $\ln(\average{Z_{m, n}^R}) - n\ln(\average{Z_m})$ does not vanish algebraically.
However, the two quantities in this difference are governed by the same saddle point at leading order. 
Consequently, the $O(V)$ contributions to $\ln\average{Z_{m,n}^R}$ and $n\ln\average{Z_m}$ cancel, yielding $\average{S_{m,n}^R(A\!:\!B)}=0$ at leading order.
This agrees with the leading-order behavior of R'enyi reflected entropy in Haar-random states \cite{Akers2022}. 
The saddle-point analysis is presented in Appendix~\ref{appendix:saddle_point_first_case_reflected_entropy}.

\subsection{Reduced density matrix of the canonical purification \label{subsec:can-pur-structure} }

Beyond the entanglement measures discussed above,
the ansatz \eqref{ergodic_tripartition} also allows us to compute the
structure of the average reduced density matrix
$\average{\rho_{AA^*}^{(m)}}$ itself.
In this subsection we present this
calculation focusing on $\rho_{AA^*}^{(m=2)}$. Our motivation is to formulate an ETH-type statement for $\rho_{AA^*}^{(2)}$ (and, analogously, for
$\rho_{AA^*}$): just as the subsystem ETH asserts that the reduced density
matrix $\rho_A$ of an individual chaotic eigenstate is close to a universal
form fixed by the eigenstate energy, we expect $\rho_{AA^*}^{(2)}$ to be
governed by a universal form to which individual eigenstates converge. The
leading-order ensemble average computed here is the candidate for that
universal form; in Sec.~\ref{sec:exact_diagonalization_multi_entropy} we combine it with exact
diagonalization to test this expectation, examining both whether the
average is reproduced and whether the fluctuations of individual
eigenstates around it are suppressed.

From the ansatz \eqref{ergodic_tripartition},
it is straightforward to write out the reduced density matrix $\rho_{AB} = \text{tr}_C(|E\rangle\langle E|)$:
\begin{equation}
    \rho_{AB} = \sum_{i,j}\sum_{i'j'}\sum_k M_{ijk}M_{i'j'k}^* |E_i^A\rangle_A |E_j^B\rangle_B \langle E_{i'}^A|_A \langle E_{j'}^B|_B.
\end{equation}
By the definition of canonical purification, 
\begin{widetext}
\begin{align}
    \nonumber |\rho_{AB}^{(2)}\rangle & = \frac{1}{\sqrt{\text{tr}(\rho_{AB}^2)}} \sum_{i,j}\sum_{i'j'}\sum_k M_{ijk}M_{i'j'k'} |E_i^A\rangle_A|E_j^B\rangle_B |E_{i'}^A\rangle_{A^*}|E_{j'}^B\rangle_{B^*}.
    \label{rho_AB_2_purification}
\end{align}
The corresponding reduced density matrix on subsystem $\mathcal{H}_A \otimes \mathcal{H}_{A^*}$ takes the form,
\begin{equation}
    \rho_{AA^*}^{(2)} = \frac{1}{\text{tr}(\rho_{AB}^2)}\sum_{i_1, i_1'} \sum_{i_2, i_2'}\sum_{j_1, j_1'}\sum_{k_1, k_2}  M_{i_1j_1k_1} M_{i_1'j_1'k_1}^* M_{i_2j_1k_2}^* M_{i_2'j_1'k_2} |E_{i_1}^A \rangle_A |E_{i_1'}^A \rangle_{A^*} \langle E_{i_2}^A |_A \langle E_{i_2'}^A |_{A^*}, 
\end{equation}
and the average value of the coefficients in the sum can be calculated as, 
\begin{align}
    & \average{M_{i_1j_1k_1} M_{i_1'j_1'k_1}^* M_{i_2j_1k_2}^* M_{i_2'j_1'k_2}} \nonumber \\ & = \average{M_{i_1j_1k_1} M_{i_2j_1k_2}^*} \ \average{M_{i_1'j_1'k_1}^* M_{i_2'j_1'k_2}} + \average{M_{i_1j_1k_1} M_{i_1'j_1'k_1}^*} \ \average{M_{i_2j_1k_2}^* M_{i_2'j_1'k_2}} \nonumber \\
    & = e^{-S(E_{i_1}^A+E_{j_1}^B+E_{k_1}^C)-S(E_{i_1'}^A+E_{j_1'}^B+E_{k_1}^C)}\delta_{i_1 i_2}\delta_{k_1 k_2}\delta_{i_1' i_2'}F(E_{i_1}+E_{j_1}+E_{k_1}-E) F(E_{i_1'}+E_{j_1'}+E_{k_1}-E) \nonumber \\ &
    \quad + e^{-S(E_{i_1}^A+E_{j_1}^B+E_{k_1}^C) - S(E_{i_2}^A+E_{j_1}^B+E_{k_2}^C)} \delta_{j_1j_1'}\delta_{i_1i_1'}\delta_{i_2i_2'}F(E_{i_1}+E_{j_1}+E_{k_1}-E) F(E_{i_2}+E_{j_2}+E_{k_1}-E).
  \label{average_coefficients_rho_AA}
\end{align}

One key difference between the average of $\rho_{AA^*}^{(2)}$ and the average of bipartite reduced density matrices is that $\rho_{AA^*}^{(2)}$ has two types of terms that dominate in different subsystem-size regimes. 
Specifically, the first term in Eq.\ \eqref{average_coefficients_rho_AA} is a diagonal coefficient corresponding to $|E_{i_1}^A \rangle_A |E_{i_1'}^A \rangle_{A^*} \langle E_{i_1}^A |_A \langle E_{i_1'}^A |_{A^*}$, and in the limit where $F(\Delta)$ is a delta function $\delta(\Delta)$ centered at $\Delta = 0$, it takes the form
\begin{align}
&
    C_{\text{diag}} (E_{i_1}^A,  E_{i_1'}^A) = 
    e^{-2S(E)} 
    \sum_{k_1} 
    e^{S_C(E_{k_1}^C) +  S_B(E - E_{i_1}^A - E_{k_1}^C) + S_B(E - E_{i_1'}^A - E_{k_1}^C)}.
    \label{diagonal_term_rho_AA}
\end{align}
On the other hand, the second term in Eq.\ \eqref{average_coefficients_rho_AA} is a ``block" coefficient corresponding to $|E_{i_1}^A \rangle_A |E_{i_1}^A \rangle_{A^*} \langle E_{i_2}^A|_A \langle E_{i_2}^A|_{A^*}$, and in the limit $F(\Delta) = \delta(\Delta)$, it takes the form
\begin{align}
&
    C_{\text{block}} (E_{i_1}^A, E_{i_2}^A) = 
   e^{-2S(E)}  
    \sum_{j_1}   
    e^{S_B(E_{j_1}^B)+S_C(E - E_{i_1}^A - E_{j_1}^B)+S_C(E - E_{i_2}^A - E_{j_1}^B)}.
    \label{block_term_rho_AA}
\end{align}
Putting these expressions together, the average reduced density matrix takes the form
\begin{align}
    & \average{\rho_{AA^*}^{(2)}} = \sum_{i_1, i_1'} C_{\text{diag}} (E_{i_1}^A, E_{i_1'}^A) |E_{i_1}^A \rangle_A |E_{i_1'}^A \rangle_{A^*} \langle E_{i_1}^A |_A \langle E_{i_1'}^A |_{A^*} 
     +  \sum_{i_1, i_2} C_{\text{block}} (E_{i_1}^A, E_{i_2}^A) |E_{i_1}^A \rangle_A |E_{i_1}^A \rangle_{A^*} \langle E_{i_2}^A|_A \langle E_{i_2}^A|_{A^*}.
    \label{eq:average_rho_AA_2_analytical}
\end{align}
\end{widetext}

\begin{figure*}[t]
    \centering
    \resizebox{0.96\textwidth}{!}{%
    \begin{tikzpicture}[>=stealth, font=\footnotesize, line width=0.6pt]
      \tikzset{site/.style={circle,draw,minimum size=17pt,inner sep=1pt,fill=white}, bond/.style={line width=1.3pt}}
      \def\dx{1.5}\def\dy{1.1}\def\panelsep{5.0}
      \newcommand{\fourcircles}{%
        \node[site] (Ai)  at (-\dx, \dy)  {$A_{\mathrm{in}}$};
        \node[site] (Asi) at ( \dx, \dy)  {$A^*_{\mathrm{in}}$};
        \node[site] (Ao)  at (-\dx,-\dy)  {$A_{\mathrm{out}}$};
        \node[site] (Aso) at ( \dx,-\dy)  {$A^*_{\mathrm{out}}$};}
      \begin{scope}[shift={(0,0)}]
        \fourcircles
        \draw[bond,blue!70!black] (Ai) -- (Ao);
        \draw[bond,blue!70!black] (Asi) -- (Aso);
        \node at (0,-\dy-0.85) {(a) diagonal ($p{=}1$), $V_B\gg V_C$};
      \end{scope}
      \begin{scope}[shift={(\panelsep,0)}]
        \fourcircles
        \draw[bond,blue!70!black,opacity=0.5] (Ai) -- (Ao);
        \draw[bond,blue!70!black,opacity=0.5] (Asi) -- (Aso);
        \draw[bond,red!75!black,opacity=0.5] (Ai) -- (Asi);
        \draw[bond,red!75!black,opacity=0.5] (Ao) -- (Aso);
        \node at (0,-\dy-0.85) {(b) mixed ($p{=}\tfrac12$), $V_B\sim V_C$};
      \end{scope}
      \begin{scope}[shift={(2*\panelsep,0)}]
        \fourcircles
        \draw[bond,red!75!black] (Ai) -- (Asi);
        \draw[bond,red!75!black] (Ao) -- (Aso);
        \node at (0,-\dy-0.85) {(c) block ($p{=}0$), $V_B\ll V_C$};
      \end{scope}
      \begin{scope}[shift={(2*\panelsep+2.5,0)}]
        \draw[bond,blue!70!black] (0,0.35)--(0.6,0.35);
        \node[right] at (0.65,0.35) {\scriptsize ket--bra bond ($A$ or $A^*$)};
        \draw[bond,red!75!black] (0,-0.25)--(0.6,-0.25);
        \node[right] at (0.65,-0.25) {\scriptsize $A$--$A^*$ entanglement};
      \end{scope}
    \end{tikzpicture}}
    \caption{
    Correlation structure of the reduced density matrix of the canonical purification, $[\rho_{AA^*}^{(2)}]$, in the $V_A/V \to 0$ limit, shown through its four indices: $A_{\mathrm{in}},A^*_{\mathrm{in}}$ (ket) and $A_{\mathrm{out}},A^*_{\mathrm{out}}$ (bra). As the ``dephasing'' parameter [see Eq.~\eqref{eq:dephasing}] is tuned from $p=1$ to $p=0$, $[\rho_{AA^*}^{(2)}]$ interpolates
    between 
    (a) a product state $\rho_A\otimes \rho_{A^*}$, where $A$ and $A^*$ are decoupled and only ket--bra bonds within each remain (diagonal regime, dominant for $V_B\gg V_C$); 
    (b) the mixed case, where the uncorrelated (product, $A$--$A^*$ decoupled) and entangled ($A$--$A^*$) channels coexist with weight $p$ and $1-p$ ($V_B\sim V_C$); and 
    (c) the thermofield double
    $\propto |\tilde{\rho}^{(2)}_A\rangle\langle\tilde{\rho}^{(2)}_A|$
    , 
    where $A$ is maximally entangled with $A^*$ (block regime, $V_B\ll V_C$). 
    }
    \label{fig:can-pur-structure}
\end{figure*}
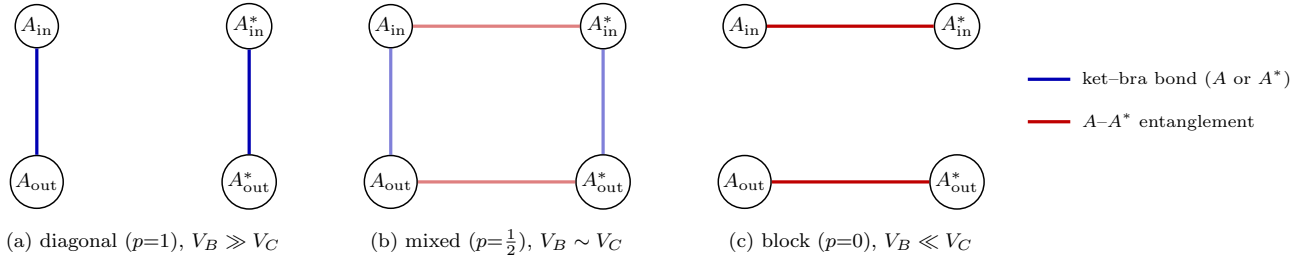

Assuming that the microcanonical entropy density is the same for all
subsystems, the density of states is larger in the subsystem with larger size. This means that $C_{\rm diag}$ dominates when the Hilbert space
dimension of subsystem $B$ is much larger than that of subsystem $C$
(i.e., $L_B\gg L_C$, where $L_K$ is the Hilbert space dimension of
subsystem $K$ and $L_K=2^{V_K}$), and $C_{\rm block}$ dominates when
$L_C\gg L_B$. When $L_B\sim L_C$, both terms are present in the
leading-order expression for the average of $\rho^{(2)}_{AA^*}$.
The block structure and the diagonal structure can be
understood diagrammatically as in Fig.~\ref{fig:can-pur-structure}; here, the ``in" and ``out" labels indicate indices for the kets and bras in the above expressions. 

\subsubsection{Structure of $\average{\rho_{AA^*}^{(2)}}$ in the $V_A/V \to 0$ limit}

To better understand the structures of $\average{\rho_{AA^*}^{(2)}}$ 
in different subsystem partitions and their relations to the phases of $(2, n)$-R\'enyi 
reflected entropy, 
we take the thermodynamic limit with $V \to \infty$ while keeping the size of $A$ 
and the energy density $E/V$ fixed, $V_A/V \to 0$. 
In this limit, the subsystem eigenstate thermalization hypothesis (ETH)~\cite{Dymarsky_2018} 
asserts that the universal form
of the reduced density matrix $\rho_A$ takes an approximate canonical form 
that is proportional to $e^{-\beta H_A}$. 
The structure of $\average{\rho_{AA^*}^{(2)}}$, by contrast, is a weighted combination of a diagonal (product) and a block (entangled) contribution whose relative weight depends on the relative sizes of $B$ and $C$. Explicitly, relegating the calculation details to Appendix \ref{appendix:vanishing_VA_rho_AA_2_calculation},
\begin{align}
    \average{\rho_{AA^*}^{(2)}} &\approx  \frac{e^{-2S(E)}}{\text{tr}(\rho_{AB})}
    \Big\{\, w_{\rm diag}\, e^{-\beta_B (H_A + H_{A^*})}
    \nonumber \\
    &\qquad 
    + w_{\rm block}\, |\tilde{\rho}_A^{(2)}\rangle\langle\tilde{\rho}_A^{(2)}|\,\Big\},
    \label{eq:rho2mix}
\end{align}
with saddle weights 
\begin{align}
&w_{\rm diag}=e^{S_C(E_{\rm diag}^*) + 2S_B(E - E_{\rm diag}^*)},
\nonumber \\
\mbox{and}\quad 
&w_{\rm block}=e^{S_B(E_{\rm block}^*) + 2S_C(E - E_{\rm block}^*)},
\end{align} 
where $E_{\rm diag}^*$ and $E_{\rm block}^*$ are the saddle-point energies 
of the diagonal and block coefficients, 
\eqref{diagonal_term_rho_AA} and\eqref{block_term_rho_AA}, respectively,
\begin{align}
|\tilde{\rho}_A^{(2)}\rangle 
= \sum_i e^{-\beta_C E_i^A}|E_i^A\rangle_A|E_i^A\rangle_{A^*}, 
\end{align}
and $\beta_B,\beta_C$ are the inverse temperatures of $B,C$ at these saddle points.
Their relative weight is captured by a single ``dephasing parameter"
\begin{equation}
  p = \frac{w_{\rm diag}}{w_{\rm diag}+w_{\rm block}} = p(L_B,L_C)\in[0,1],
  \label{eq:dephasing}
\end{equation}
which depends only on the subsystem sizes.

In the two limits $L_B\ll L_C$ and $L_B\gg L_C$,
Eq.\ \eqref{eq:rho2mix} reduces to 
the simpler forms.
For $L_B \ll L_C$ the block term dominates ($p\to0$) and the average is the pure state,
\begin{equation}
     \average{\rho_{AA^*}^{(2)}} \propto  |\tilde{\rho}_A^{(2)}\rangle\langle\tilde{\rho}_A^{(2)}|
    \qquad (L_B \ll L_C),
    \label{eq:rho2tfd}
\end{equation}
where $A$ and $B$ are both nearly maximally entangled with $C$, 
leaving little $A$-$B$ entanglement; indeed, the leading $(2,n)$-R\'enyi reflected entropy vanishes here.
For $L_B \gg L_C$ the diagonal term dominates ($p\to1$) and,
\begin{equation}
    \average{\rho_{AA^*}^{(2)}} \propto e^{-\beta_B(H_A + H_{A^*})}
    \qquad (L_B \gg L_C),
    \label{eq:rho2dec}
\end{equation}
which is a product of two thermal density matrices at temperature $\beta$, with one copy in the original Hilbert space and the other in the purifying space;
at $\beta=0$ the leading $(2,n)$-R\'enyi reflected entropy scales as $2V_A$ here. 
As the dephasing parameter $p$ is varied, $\average{\rho^{(2)}_{AA^*}}$
interpolates continuously between the pure, block-dominated form
\eqref{eq:rho2tfd} ($p\to0$), the product, diagonal-dominated form
\eqref{eq:rho2dec} ($p\to1$), and the balanced combination \eqref{eq:rho2mix}
at $p=1/2$.

\subsubsection{Connections to the Thermofield and Thermo-Mixed Doubles} \label{sec:tfd-tmd}

Intriguingly, as we now show, the form of the reduced density matrix for the above three values of the dephasing parameter at $p=0,1/2,1$ bear a close relationship to the three reference states introduced in Refs.~\onlinecite{Verlinde2020,Verlinde2021}---a product state, the thermo-\emph{mixed} double, and the thermofield double---in the seemingly distinct context of describing the two-sided eternal black hole in gravitational physics.
This connection can be first seen by computing the mutual information between $A$ and $A^*$, which we turn to next.
In terms of the $m=2$ one-sided reduction,
\begin{equation}
  \rho_A^{(2)}\;\equiv\;\mathrm{tr}_{A^*}\rho_{AA^*}^{(2)},
  \label{eq:m2_rhoA}
\end{equation}
the relevant mutual information is,
\begin{align}
  I^{(2)}(A\!:\!A^*)
  &\;=\;S_{\rm vN}(\rho_A^{(2)})+S_{\rm vN}(\rho_{A^*}^{(2)})-S_{\rm vN}(\rho_{AA^*}^{(2)}) \nonumber\\
  &\;=\;2\,S_{\rm vN}(\rho_A^{(2)})-S_{\rm vN}(\rho_{AA^*}^{(2)}),
  \label{eq:m2_def}
\end{align} 
where the second equality uses the mirror symmetry
$S_{\rm vN}(\rho_{A^*}^{(2)})=S_{\rm vN}(\rho_A^{(2)})$ of the canonical
purification. The quantity $S_{\rm vN}(\rho_{AA^*}^{(2)})$ is the $m=2$
reflected entropy,
i.e., the $n\to1$ limit of the $(2,n)$-R\'enyi reflected
entropy.

The reflected entropy follows from the dephasing structure of
\eqref{eq:rho2mix}: writing $\rho_{AA^*}^{(2)}$ as the weighted combination of
a pure (entangled) and a product piece with weight $p$ \eqref{eq:dephasing},
its von Neumann entropy is, to leading order,
\begin{equation}
  S_{\rm vN}(\rho_{AA^*}^{(2)}) \sim 2p\,S_{\rm vN}(\rho_A^{(2)}) + H(p),
  \label{eq:m2_SR_p}
\end{equation}
with $H(p)=-p\ln p-(1-p)\ln(1-p)=O(1)$ the binary entropy, so that
\eqref{eq:m2_def} gives,
\begin{equation}
  I^{(2)}(A\!:\!A^*) \sim 2\,S_{\rm vN}(\rho_A^{(2)})\,(1-p).
  \label{eq:m2_I_p}
\end{equation}
As $p=p(L_B,L_C)$ sweeps from zero to unity with the partition size, $I^{(2)}$
interpolates continuously from $2S_{\rm vN}(\rho_A^{(2)})$ to zero, passing
through $S_{\rm vN}(\rho_A^{(2)})$ at $p=1/2$, so that,
\begin{equation}
  \frac{I^{(2)}(A\!:\!A^*)}{S_{\rm vN}(\rho_A^{(2)})}\;\to\;2,1,0
  \quad (p=0,\tfrac12,1).
  \label{eq:m2_trich}
\end{equation}
The value of zero at $p=1$ is protected at all temperatures, since a product
state has vanishing mutual information.

These values match those of the three reference states discussed in
Refs.~\cite{Verlinde2020,Verlinde2021}. There, a one--sided thermal state
$\rho_{\rm th}=e^{-\beta H}/Z=\sum_n p_n|n\rangle\langle n|$ (with
$p_n=e^{-\beta E_n}/Z$) has canonical purification given by the thermofield double,
$|{\rm TFD}\rangle=\sum_n\sqrt{p_n}\,|n\rangle_L|n\rangle_R$, a pure $L$-$R$
entangled state whose one--sided reduction
$\mathrm{tr}_R|{\rm TFD}\rangle\langle{\rm TFD}|=\rho_{\rm th}$ is thermal. 
The thermo-mixed double (TMD) has the same thermal marginal but only classical
$L$-$R$ correlations, obtained by dephasing the TFD in the energy basis,
\begin{equation}
  \rho_{\rm TMD}=\sum_n p_n\,|n\rangle_L\langle n|\otimes|n\rangle_R\langle n|,
  \label{eq:TMD}
\end{equation}
while the decoupled product state is $\rho_L\otimes\rho_R$. All three share the
one-sided entropy $S_{\rm BH}\equiv-\sum_n p_n\ln p_n$ and are distinguished
by the $L$-$R$ mutual information $I_{LR}=S_L+S_R-S_{LR}$, equal to
$2S_{\rm BH}$, $S_{\rm BH}$ and $0$ for the TFD, TMD and decoupled state.\footnote{Here the subscript $\mathrm{BH}$ is used in analogy with the context of Refs.~\cite{Verlinde2020,Verlinde2021}, where each side corresponds to one of the black holes of the eternal black hole spacetime geometry. }

Under the identification $A\leftrightarrow L$, $A^*\leftrightarrow R$, the
block term \eqref{eq:rho2tfd} carries $L$-$R$ entanglement as in the TFD, the
diagonal term \eqref{eq:rho2dec} is an uncorrelated product as in the decoupled
state, and the balanced combination \eqref{eq:rho2mix} plays the role of the
TMD: while its precise matrix elements do not coincide with those of
$\rho_{\rm TMD}$ \eqref{eq:TMD}, its mutual information does. The
trichotomy \eqref{eq:m2_trich} thus reproduces $I_{LR}$ with
$S_{\rm BH}\to S_{\rm vN}(\rho_A^{(2)})$, and $\average{\rho^{(2)}_{AA^*}}$
realizes the entire TFD-TMD-decoupled interpolation at any $\beta$;
the three reference states being its $p=0,\tfrac12,1$ slices, with the dephasing
parameter set by the subsystem-size ratio.

These forms nonetheless fail to reproduce one feature of the reference states.
The reference states of \cite{Verlinde2020,Verlinde2021} are built from a
\emph{single} thermal state and so share the same one-sided entropy $S_{\rm BH}$,
whereas the one-sided entropy $S_{\rm vN}(\rho_A^{(2)})$ differs between the
regimes at $\beta\neq0$:
\begin{equation}
  S_{\rm vN}(\rho_A^{(2)})\big|_{\rm block}=S_A(2\beta)\;\neq\;
  S_A(\beta)=S_{\rm vN}(\rho_A^{(2)})\big|_{\rm diag},
  \label{eq:m2_noncommon}
\end{equation}
with the balanced form \eqref{eq:rho2mix} in between; here
$S_A(\beta)\equiv S_{\rm vN}(e^{-\beta H_A}/Z_A)$ is the thermal entropy of $A$
at inverse temperature $\beta$. This is because the
one-sided reduction $\rho_A^{(2)}=\mathrm{tr}_B\rho_{AB}^{(2)}$ is built from
the \emph{squared} state $\rho_{AB}^2$, not the physical reduced density
matrix, so its effective temperature is $2\beta$ in the block regime
\eqref{eq:rho2tfd} but $\beta$ is in the diagonal regime \eqref{eq:rho2dec} (with
$\beta_B\simeq\beta_C\simeq\beta$). At
$\beta=0$, by contrast, $S_A(2\beta)=S_A(\beta)=V_A\ln2$, so all three regimes
share the common $S_{\rm BH}=V_A\ln2$ and the reference structure is realized
exactly.

This regime dependence is an artifact of working with $m=2$. For the true
($m=1$) reflected entropy the one-sided reduction is the physical reduced
density matrix, $\mathrm{tr}_{A^*}\rho_{AA^*}=\mathrm{tr}_B\rho_{AB}=\rho_A$,
which is fixed by the global energy density and is therefore the \emph{same}
thermal state in all three regimes. The three regimes then share a single
one-sided thermal state $\rho_A$ at every temperature --- precisely the
structure of the reference states \cite{Verlinde2020,Verlinde2021}, which are
built from a single thermal state --- whereas the $m=2$ object shares one only
at $\beta=0$, where $\rho_A^{(2)}\to\rho_A$ and $m=1$ and $m=2$ coincide.
We therefore expect $\rho_{AA^*}$ to, at least qualitatively, interpolate between these three reference states, though we do not attempt to establish this quantitatively.

\section{Tests of Analytical Predictions via Exact Diagonalization} 

\label{sec:exact_diagonalization_multi_entropy}

We now ask how well the energy-constrained Haar random ansatz
\eqref{ergodic_tripartition} describes eigenstates in tripartite quantum
chaotic systems. To this end, we compare its analytical predictions for
tripartite entanglement probes, obtained in Sec.\ \ref{section_2}, against
exact diagonalization of a chaotic Hamiltonian. 
Specifically, we examine the mixed-field Ising model (MFIM), which has been shown to exhibit quantum chaotic behavior under multiple diagnostics \cite{Kim_2013, Zhang_2015, Roberts_2015, Khemani_2018}. 
The Hamiltonian is given by
\begin{equation}
    H = -\sum_{k=1}^{n-1} \sigma^z_{k} \otimes \sigma^z_{k+1} - g\sum_{k=1}^n \sigma^x_k - h_1\sigma^z_1 - h_n\sigma^z_n - h\sum_{k=2}^{n-1}\sigma_k^z.
    \label{mixed_field_Ising_model}
\end{equation}
Here, $\sigma^{x, z}_i$ are the Pauli matrices on site $i$. 
We study systems with different transverse and longitudinal fields $g$ and $h$. 
In order to avoid the effect of additional symmetry constraints on entanglement, we adopt open boundary conditions and choose $h_1 \neq h_n \neq h$ to break the translation and inversion symmetry. As a result, energy is the only conserved quantity in the system.

Throughout this section,
we focus on the R\'enyi index $n=2$. Recall that, at this index, the R\'enyi-$2$ multi-entropy coincides with the $(2, 2)$-R\'enyi reflected entropy up to a R\'enyi-$2$ bipartite entropy term [Eq.\ \eqref{multi_entropy_definition}]: $Z_{2,2}^R = Z_2^{(3)}$. 
The $n=2$ case is numerically convenient
and enough to capture the leading structure of tripartite entanglement probed in our analysis.

To reduce finite size effects, when computing $S_2^{(3)}$ from 
the analytical predictions, we use $\average{Z_{2, \text{exact}}^{(3)}}$ and include all orders in the expression. 
The full expression for $\average{Z_{2, \text{exact}}^{(3)}}$ is provided in Appendix \ref{appendix:full_expression_Z_2_3}. 
We note that the difference between the full and the leading order expressions of $\average{Z_{2, \text{exact}}^{(3)}}$ scales as $O(e^{-\alpha V})$ and is negligible in the thermodynamic limit.

Moreover, when implementing the energy conservation constraint, we allow the window function to have a finite width $\Delta E$: the window function $F(E^A_i + E^B_j + E^C_k - E)$ is set equal to one when the difference between the subsystem energy sums and the total energy lies
within $\Delta E$, and zero otherwise. 
In Appendix \ref{appendix:check_on_delta_E}, we explicitly verify that $\average{S_{2}^{(3)}}$ is insensitive to $\Delta E$ over a reasonable range. 
The density of states used in our computation is obtained directly from exact diagonalization of the subsystem Hamiltonians $H_A$, $H_B$ and $H_C$ corresponding to the partition we used.

\begin{figure*}[t]
    \centering
    \includegraphics[width=1\textwidth]{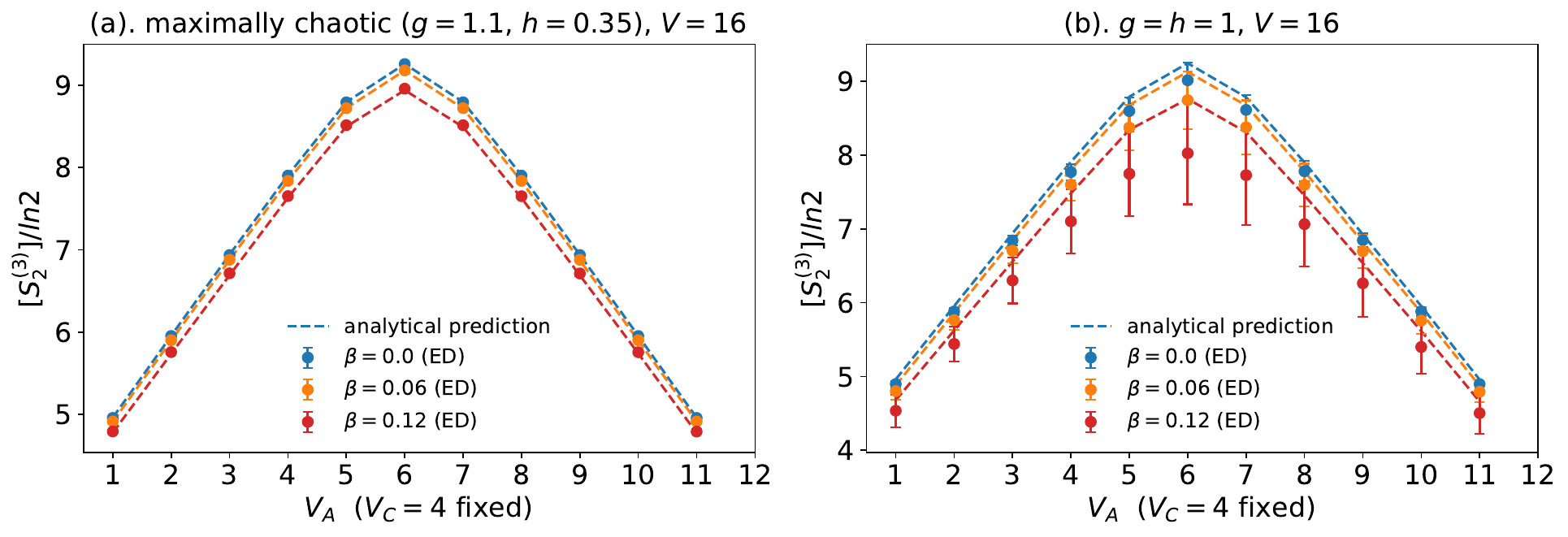}
    \caption{Comparison between $S_2^{(3)}$ obtained from exact diagonalization (circular markers) and the analytical prediction (dashed lines) for energy eigenstates of the mixed-field Ising model
    for (a) maximally chaotic parameters $(g,h)=(1.1,0.35)$ and (b) $(g,h) = (1,1)$. We consider three inverse temperatures $\beta = 0, 0.1, 0.2$. We fix the total system size $V = 16$ and $C$ subsystem size $V_C = 4$ while varying $V_A$. Eigenstate-to-eigenstate fluctuations are shown 
    as error bars. The dashed lines are the analytical predictions at each 
    $\beta$. 
    The exact-diagonalization averages decrease with $\beta$ and lie below the the analytical predictions;
    the fluctuations are small in (a) 
    but substantially larger in (b), 
    and grow with $\beta$.
    }
    \label{fig:S_2_3_fluctuations}
\end{figure*}

\subsection{Tripartite multi-entropy: Comparison with analytical predictions} 
\label{comparison_analytical_exact_diagonalization}

\begin{figure*}[t]
    \centering
    \includegraphics[width=0.8\textwidth]{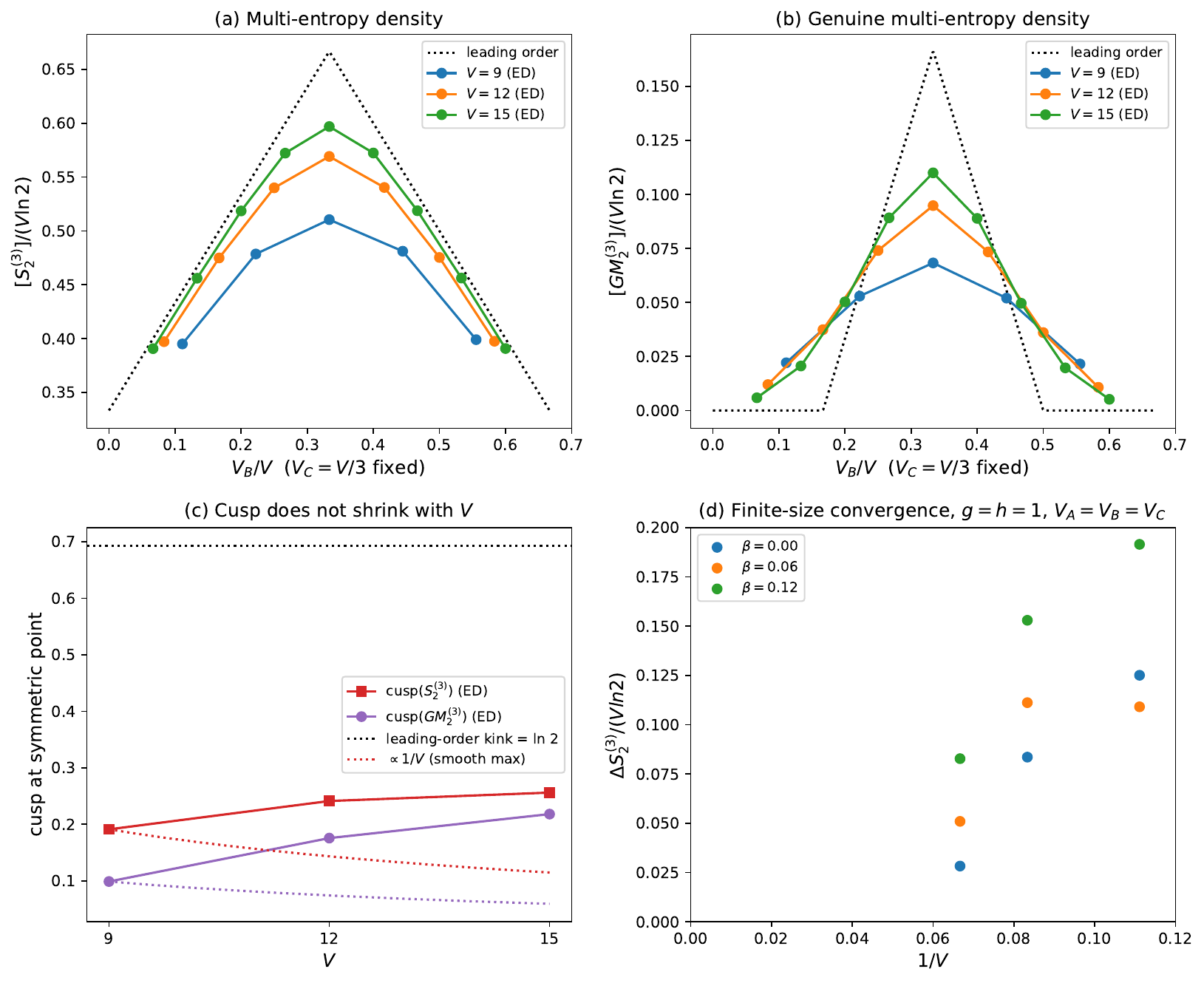}
    \caption{(a) Intensive R\'enyi-2 multi-entropy $S_2^{(3)}/(V\ln 2)$ and (b) genuine multi-entropy $GM_2^{(3)}/(V\ln 2)$ along the cut through the symmetric tripartition ($V_C = V/3$ fixed, $V_B$ varied, $V_A = V - V_B - V_C$) for the maximally chaotic [i.e. $(g,h)=(1.1,0.35)$] mixed-field Ising model at $\beta \approx 0$, with system sizes $V = 9, 12, 15$. 
    The dotted line is the leading-order (large-$V$) prediction $S_2^{(3)}/(V\ln 2) = 1 - \max(f_A, f_B, f_C)$, $f_K = V_K/V$, a single universal curve with a kink at the symmetric point $V_B/V = 1/3$. 
   (c) The cusps of the R\'enyi multi-entropy and of the genuine multi-entropy at the symmetric point, defined as $S(V_B{=}V/3) - \tfrac{1}{2}[S(V_B{=}V/3-1) + S(V_B{=}V/3+1)]$ with $S = S_2^{(3)}$ or $GM_2^{(3)}$ along the cut of panels (a,b), do not shrink with $V$ and lie below the leading-order kink value $\ln 2$, consistent with a genuine non-analyticity rounded only over a finite-size window.
    (d) Finite-size convergence of the deviation $\Delta S_2^{(3)}/V = |\average{S_2^{(3)}}_{\rm ED} - \average{S_2^{(3)}}_{\rm leading}|/V$ between exact diagonalization and the leading-order prediction at the symmetric tripartition $V_A = V_B = V_C$, for $g=h=1$ and different values of $\beta$;
    the deviation decreases with system size, suggesting that the residual discrepancy
    in Fig.~\ref{fig:S_2_3_fluctuations}(b) is a finite-size effect. 
    } 
    \label{fig:multi_entropy_cusp_draft}
\end{figure*}

We begin by comparing the analytical predictions for the multi-entropy with exact diagonalization results for two representative parameter sets: 
one with $(g,h)=(1.1, 0.35)$
in the maximally chaotic regime identified in Ref.~\cite{Rodriguez_Nieva_2024}, and the other with $(g, h) = (1,1)$ corresponding to the parameter set studied in Ref.~\cite{Lu_2017} 
in the context of bipartite entanglement in chaotic eigenstates. 
In both cases, we choose 
$(h_1, h_n) = (-0.25,0.25)$ 
and fix the system size to be $V = 16$.

Since our analytical predictions are expected to hold not only at infinite temperature (i.e. in the middle of the spectrum) but also at high and finite temperatures, we examine the consistency between the analytical and numerical results for $S_2^{(3)}$ at multiple inverse temperatures $\beta$.
For a chaotic Hamiltonian $H$, 
the inverse temperature $\beta$ 
associated with eigenstate energy $E$ is determined by matching $E$ with the mean energy of the corresponding thermal ensemble,
$\langle H \rangle_{\text{th}}(\beta) = \text{tr}(H e^{-\beta H})/\text{tr}(e^{-\beta H}) = E$.
In the mixed-field Ising model, $\beta = 0$ (infinite temperature) 
corresponds to $E = 0$ at the middle of the spectrum, while nonzero $\beta$ corresponds to moving away from the spectral center.
We parametrize our results by $\beta$ rather than $E$ because it provides a more natural way to compare across different system sizes. 
For a local many-body Hamiltonian, fixing $\beta$ fixes the energy density, while the total energy scales with system size.

The comparisons between the analytical predictions and exact diagonalization results are shown in Fig.\ \ref{fig:S_2_3_fluctuations}. 
We find excellent agreement in the maximally chaotic case for all four values of $\beta$ probed in our analysis. 
In the case $g = h = 1$, the analytical predictions remain reasonably consistent with the exact diagonalization results. The deviations can be attributed to finite size effects (see Fig.\ \ref{fig:multi_entropy_cusp_draft}).  
Additionally, the spread of $S_2^{(3)}$ across the eigenstates in the energy window
provides a finer diagnostic than the mean alone: 
in the maximally chaotic case the eigenstate-to-eigenstate standard deviation is small and only weakly temperature-dependent, whereas for $g=h=1$ it is markedly larger and grows with $\beta$, consistent with this parameter set lying farther from the maximally chaotic regime.

The leading-order multi-entropy is piecewise linear in the subsystem sizes, $\average{S_2^{(3)}}/(V\ln 2) = 1 - \max(f_A, f_B, f_C)$ with $f_K = V_K/V$, and develops a kink at the symmetric tripartition $V_A = V_B = V_C$, where the three saddles $V_A + V_B$, $V_A + V_C$ and $V_B + V_C$ become degenerate.
Figure~\ref{fig:multi_entropy_cusp_draft} examines this competing-saddle region directly, along the cut $V_A = V/3$.
Plotted as an intensive density, the leading-order predictions for different system sizes collapse onto a single universal curve (dotted), and the exact diagonalization results approach it from below as $V$ increases.
The cusp at the symmetric point---the height of the kink relative to its nearest-neighbor partitions---does not shrink as $1/V$, the behavior expected of a smooth maximum; instead it stays far above the $1/V$ reference and below the leading-order value of $\ln 2$.
This indicates that the kink is a genuine non-analyticity of the multi-entropy, rounded only over a finite-size window at the cusp, rather than a finite-size artifact.
The same feature appears, more sharply, in the genuine multi-entropy $GM_2^{(3)}$.

\subsection{Breakdown of the ansatz: Tripartite versus bipartite multi-entropy}
\label{section: comparison_bipartite_tripartite}

\begin{figure*}[t]
    \includegraphics[width=1\textwidth]{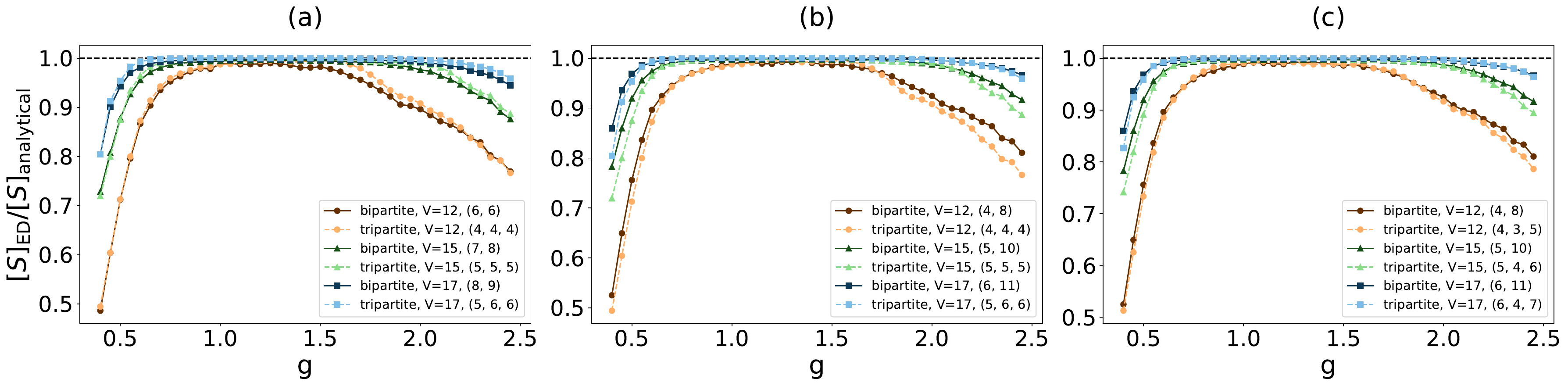}
    \caption{
   Comparison between the ratios
$\average{S_2^{(3)}}_{\rm ED}/\average{S_2^{(3)}}_{\rm analytical}$
(tripartite) and
$\average{S_2}_{\rm ED}/\average{S_2}_{\rm analytical}$
(bipartite)
as functions of the transverse field $g$
for several system sizes, with longitudinal field fixed at $h = 0.35$.
For all panels, exact-diagonalization results are averaged over
a microcanonical window centered at the energy corresponding to
$\beta=0.05$.
(a) Bipartite and tripartite partitions are chosen near the
corresponding entanglement-transition regimes of the constrained
Haar-random prediction, providing a sensitive test of the theory.
(b) The same tripartite partitions as in (a) are used, while the
bipartite partition is obtained by grouping two of the three
tripartite subsystems together. 
(c) The same bipartite partitions as in (b), but with tripartite
partitions chosen away from the entanglement-transition regime,
allowing us to assess the dependence on subsystem partition. 
In each figure, the bipartite and tripartite curves are labeled by the partitions $(V_A, V_B)$ and $(V_A, V_B, V_C)$ respectively.
}
    \label{fig:bipartite_tripartite_compare}
\end{figure*}

While the preceding discussion demonstrated agreement between exact diagonalization results and the energy-constrained Haar-random predictions at representative points in parameter space, we now investigate how broadly this agreement persists as the Hamiltonian parameters are varied. 
In particular, we ask whether tripartite entanglement deviates from the constrained Haar-random prediction earlier than bipartite entanglement. As we will show below, we find no evidence for such an earlier breakdown.

In Fig.\ \ref{fig:bipartite_tripartite_compare},
we show the ratios between exact diagonalization and constrained Haar-random predictions for both the tripartite entropy $S_2^{(3)}$ and the bipartite entropy $S_2$. 
For all subsystem partitions considered, $S_2^{(3)}$ exhibits a broad parameter regime in which the ratio remains close to unity, demonstrating
that the constrained Haar-random ansatz
properly captures 
tripartite entanglement over an extended region of parameter space.
Moreover, the region of good agreement broadens with increasing system size.
In regions where the deviations between exact diagonalization and the analytical prediction remain appreciable, the deviations are reduced as the system size increases.

A key observation is that the parameter regime over which 
the constrained Haar-random ansatz accurately describes $S_2^{(3)}$ closely tracks the corresponding range of $S_2$. This agreement persists across the subsystem partitions and temperatures examined. We therefore find no evidence that genuinely tripartite correlations lead to an earlier breakdown of the constrained Haar-random description than that observed for bipartite entanglement.
Outside the region of good agreement, however, the finite-size corrections to $S_2$ and $S_2^{(3)}$ differ quantitatively and depend on the subsystem partition, suggesting distinct finite-size scaling behaviors.

\begin{figure}[htbp]
    \includegraphics[width=0.4\textwidth]{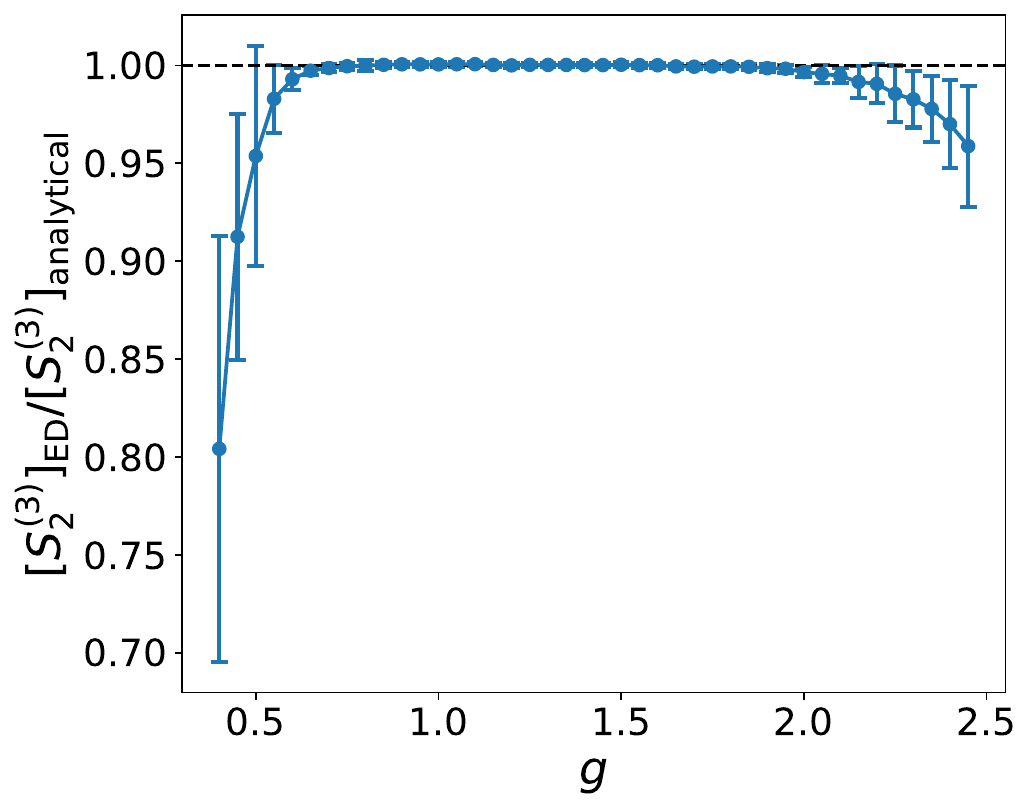}
    \caption{$\average{S_2^{(3)}}_{\text{ED}}/\average{S_2^{(3)}}_{\text{analytical}}$ as a function of transverse field $g$ with fixed $h = 0.35$. We take $V = 17$ with $V_A = 5$ and $V_B = V_C = 6$, while we set the inverse temperature to be $\beta = 0.05$. The error bars are given by the standard deviations normalized by the analytical prediction: $\sigma(S_2^{(3)})/\average{S_2^{(3)}}_{\text{analytical}}$.}
    \label{fig:standard_deviation_V_17}
\end{figure}

We also briefly discuss the fluctuations of
$S_2^{(3)}$ across eigenstates.
An example is shown in Fig.~\ref{fig:standard_deviation_V_17},
where we plot both
$\average{S_2^{(3)}}_{\rm ED}/
\average{S_2^{(3)}}_{\rm analytical}$
and the corresponding standard deviation.
As expected, fluctuations become appreciable in parameter
regimes where the average value deviates significantly from
the constrained Haar-random prediction.
By contrast, throughout the region of good agreement,
the standard deviation remains very small,
providing evidence that the R\'enyi-2 multi-entropy is
typical among chaotic eigenstates.
The fluctuations of $S_2^{(3)}$ and of the associated reduced
density matrix $\rho^{(2)}_{AA^*}$ will be investigated in
greater detail in the next subsection.

\subsection{Structure of $\rho_{AA^*}^{(2)}$ and $\rho_{AA^*}$: Evidence for a tripartite subsystem ETH generalization}
\label{rho_AA}

We now move to a stronger check on the constrained Haar-state ansatz by investigating the validity of the prediction of Eq.~\eqref{eq:average_rho_AA_2_analytical} for the structure of the reduced density matrix of the canonical purification, $\rho_{AA^*}^{(2)}$. We again focus on the R\'enyi index $n=2$ and briefly discuss the von Neumann limit of $\rho_{AA^*}$. 
If this prediction 
holds in the MFIM, this would provide evidence that the structure of $\rho_{AA^*}^{(2)}$ in a chaotic eigenstate follows a \emph{universal} form set by the energy. In this sense, the prediction of Eq.~\eqref{eq:average_rho_AA_2_analytical} may be understood as a tripartite generalization of the \emph{subsystem} ETH ansatz \cite{Dymarsky_2018}, as we now describe.

Indeed, the relation of $\rho_{AA^*}^{(2)}$ and $\rho_{AA^*}$ to the (R\'enyi) reflected entropies is analogous to that of 
the reduced density matrix $\rho_A$ to the (R\'enyi) bipartite entropies. 
In chaotic systems, there is now substantial evidence that the structure of $\rho_A$ takes a universal form, as dictated by the subsystem ETH ansatz
\cite{Dymarsky_2018}.
One of the central postulates of this ansatz is
that for an eigenstate with energy $|E_i\rangle$ and a subsystem $A$ smaller than half of the total system $(V_A < V_{\bar{A}})$, the reduced density matrix $\rho_A^i = \text{tr}_{\bar{A}} |E_i\rangle \langle E_i|$ is exponentially close to some universal density matrix $\rho_A (E = E_i)$ that depends smoothly on the eigenstate energy:
\begin{equation}
    D(\rho_A^i, \rho_A(E = E_i)) \sim O(e^{V_A - S(E_i)/2}).
    \label{subsystem_ETH}
\end{equation}
Here, the distance between two matrices is  measured by the trace distance, 
\begin{equation}
    D(\rho, \sigma) = \frac{1}{2} \lvert\lvert \rho - \sigma \rvert\rvert_1 = \frac{1}{2} \text{tr}\left( \sqrt{(\rho-\sigma)^\dagger (\rho - \sigma)} \right).
    \label{trace_distance_definiton}
\end{equation}

The implication of Eq.\ \eqref{subsystem_ETH} can be understood in two steps.  
First, Eq.\ \eqref{subsystem_ETH} implies that in an ensemble of eigenstates 
within a small window $[E_i - \Delta E, E_i + \Delta E]$ around a given energy $E_i$, the fluctuations of the individual reduced density matrices around their average are exponentially suppressed. 
Second, Eq.\ \eqref{subsystem_ETH} further states that there is a universal form of reduced density matrices that depends only on eigenstate energy. 
Specifically, when $V \to \infty$ and $V_A/V$ are finite and fixed, this universal form $\rho_A(E)$ can be described by a semiclassical expression consistent with the ergodic bipartition ansatz \cite{Dymarsky_2018, Lu_2017}. 
When $V_A/V \to 0$, on the other hand, $\rho_A(E)$ takes the thermal, or canonical, form,
$\text{tr}_{\bar{A}} (e^{-\beta H})$, or approximately $e^{-\beta H_A}$ if we neglect near-boundary correlations \cite{Dymarsky_2018, Garrison_2018, Goldstein_2006, Popescu_2006}. Here, $\beta$ is the inverse temperature determined by the eigenstate energy, and $H_A$ is the restriction of $H$ to subsystem $A$. This canonical form can also be derived from the ergodic bipartition ansatz by taking the appropriate limit $E_i^A \ll E_i$.

As in the original presentation of ETH, the subsystem ETH ansatz implies that expectation values of operators supported on a subsystem smaller than half of the total system evaluate to thermal expectation values at an effective temperature set by the energy density. 
However, the subsystem ETH ansatz is stronger, in that it also implies that nonlocal measures, like the bipartite R\'enyi entanglement entropies, also exhibit universal behavior controlled by the structure of the reduced density matrix, which is set by the effective temperature \cite{Dymarsky_2018}. It is with this in mind that we wish to provide evidence for the analogous claim that $\rho_{AA^*}^{(2)}$ (and $\rho_{AA^*}$) in a chaotic eigenstate follow a universal form, set by Eq.~\eqref{eq:average_rho_AA_2_analytical}. As such, the following may be viewed as evidence for a \emph{tripartite} generalization of the subsystem ETH ansatz.

As in the discussion of the standard subsystem ETH ansatz,
we proceed in two steps.
First, we examine whether, in a narrow energy window, the $\rho_{AA^*}$ and $\rho_{AA^*}^{(2)}$ of individual eigenstates within that window are close to their energy-window averages.
We then study the leading-order structure of the average of $\rho_{AA^*}^{(2)}$, whose analytical value can be computed directly from
the tripartition ansatz  Eq.~\eqref{ergodic_tripartition}, as described in Sec.\ \ref{subsec:can-pur-structure}. 

\subsubsection{Fluctuations around the ensemble average}
\begin{figure*}[htbp]
    \centering    
    \includegraphics[width=1 \textwidth]{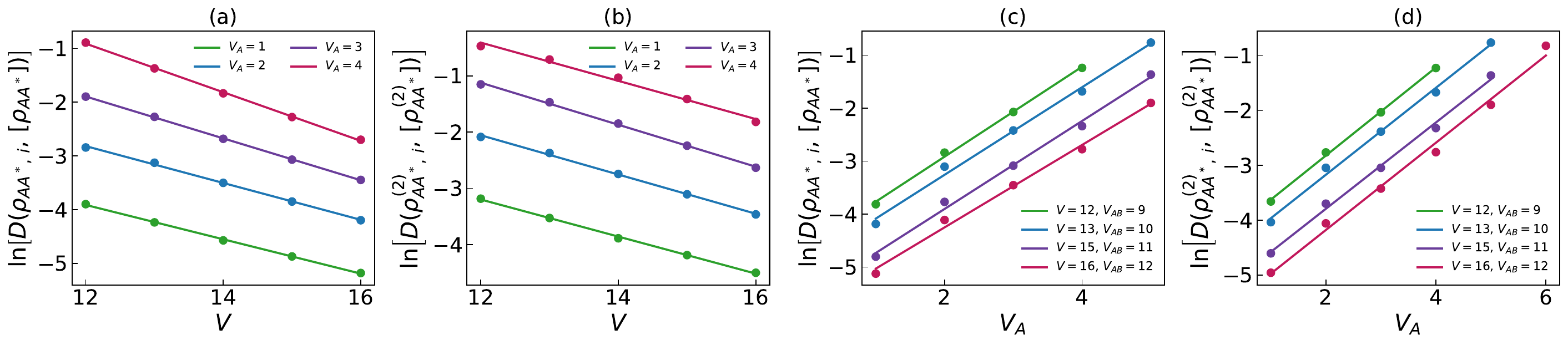}
    \caption{
    (a,c) Average trace distance between $\rho_{AA^*}$ for an eigenstate in an energy window $\Delta E$ and the average $[\rho_{AA^*}]$ of all eigenstates within that window in the mixed-field Ising model with $(g,h) = (1.1,0.35)$. (b,d) The same plots as in (a,c), but for $\rho_{AA^*}^{(2)}$.
    The average is taken over the energy window $\Delta E = 0.15$ around the middle of the spectrum $E = 0$. 
    We take (a, b) $V_B =2$ fixed
    and (c, d) $V_{AB}/V \approx 0.75$.
    }     \label{fig:trace_distance_individual_avg_large_VC}
\end{figure*}

Following the same logic as for the subsystem ETH ansatz for $\rho_A$,
the first step in understanding the structure of $\rho_{AA^*}$ and $\rho_{AA^*}^{(2)}$ is to examine their fluctuations within a narrow energy window. More precisely, we wish to see how strongly the $\rho_{AA^*}$ and $\rho_{AA^*}^{(2)}$ for individual eigenstates in this window deviate from the ensemble average. If such fluctuations are sufficiently suppressed, then an analytical description of the ensemble average can also serve as an effective description of individual eigenstates. In particular, this would provide evidence that $\rho_{AA^*}$ and $\rho_{AA^*}^{(2)}$ of each individual eigenstate follow a universal structure, in the spirit of subsystem ETH.

To quantify the fluctuations, we consider an ensemble of eigenstates with energies in the range $[E-\Delta E,E+\Delta E]$ around a fixed energy $E$ near the middle of the spectrum. We measure the average trace distance between an individual density matrix and the corresponding ensemble average:
\begin{equation}
    \average{D(\rho_{AA^*,i}^{(n)},
    \average{\rho_{AA^*}^{(n)}})}
    =
    \frac{1}{N}
    \sum_{i=1}^{N}
    D(\rho_{AA^*,i}^{(n)},\average{\rho_{AA^*}^{(n)}}),
\end{equation}
with $n=1,2$, where $N$ denotes the number of eigenstates in the energy window.

There are two regimes for us to consider, corresponding to when $AB$ is smaller than or larger than half the total system size; the original subsystem ETH ansatz does apply to $\rho_{AB}$ in the former case while it does not in the latter.
We first consider the regime $V_{AB}<V/2$. In this regime, the subsystem ETH ansatz predicts that fluctuations of $\rho_{AB}$ around its ensemble average are exponentially suppressed \cite{Dymarsky_2018}. Since $\rho_{AA^*}$ and $\rho_{AA^*}^{(2)}$ are constructed from $\rho_{AB}$, it is natural to expect similar behavior. This expectation is supported by the numerical results shown in Fig.\ \ref{fig:trace_distance_individual_avg_large_VC}. When $V_A$ and $V_B$ are held fixed and the total system size is increased, the average trace distance decreases approximately exponentially with $V$, independent of whether $V_A<V_B$ or $V_A>V_B$.

More interesting behavior emerges when $V_{AB}>V/2$. In this regime, subsystem ETH no longer applies and hence \emph{cannot} be used to deduce the suppression of fluctuations in $\rho_{AB}$ itself. Nevertheless, we find that fluctuations of $\rho_{AA^*}$ and $\rho_{AA^*}^{(2)}$ remain strongly suppressed. To investigate this regime, we keep $V_{AB}/V \approx 0.75$ fixed while varying the system size and subsystem partition. As shown in Fig.\ \ref{fig:trace_distance_individual_avg_large_VC}, the average trace distance decreases significantly as $V_B-V_A$ increases. In particular, fluctuations remain small when $V_A \ll V_B$, despite the absence of a corresponding suppression mechanism for $\rho_{AB}$ from subsystem ETH.

A natural interpretation is that the large $BB^*$ sector in the purified Hilbert space effectively washes out fluctuations when traced over. That is, while region $C$ is not sufficiently large to serve as a bath for $AB$, the region $B$ is large enough for $BB^*$ to serve as a bath for $AA^*$ in the purified system. We also observe evidence that the fluctuations continue to decrease with increasing system size when subsystem ratios are held fixed.

Overall, our results provide strong evidence that fluctuations of $\rho_{AA^*}$ and $\rho_{AA^*}^{(2)}$ are suppressed whenever $V_A$ is much smaller than \emph{either} $V_B$ or $V_C$. 
In these regimes, the ensemble-averaged density matrices studied in the next subsection are therefore expected to provide an accurate description of typical individual eigenstates as well. We also note that, analogous to the suppression of fluctuations in $\rho_{\bar A}$ when $V_A>V_{\bar A}$ in the bipartite setting, one expects fluctuations of $\rho_{BB^*}$ to be suppressed when $V_A$ is the largest subsystem in the tripartition.

Finally, while the data are broadly consistent with an approximately exponential suppression of fluctuations, the precise scaling behavior remains unclear. In particular, the observed slopes vary somewhat across subsystem partitions, and it is not yet possible to determine whether these variations arise from finite-size effects or from a more intricate dependence on subsystem ratios. We leave a detailed analysis of the scaling form of
$
\average{D(\rho_{AA^*,i}^{(n)},\average{\rho_{AA^*}^{(n)}})}$
to future work.

\subsubsection{Comparison with analytical predictions}
\begin{figure*}[htbp]
    \centering
    \includegraphics[width=1\textwidth]{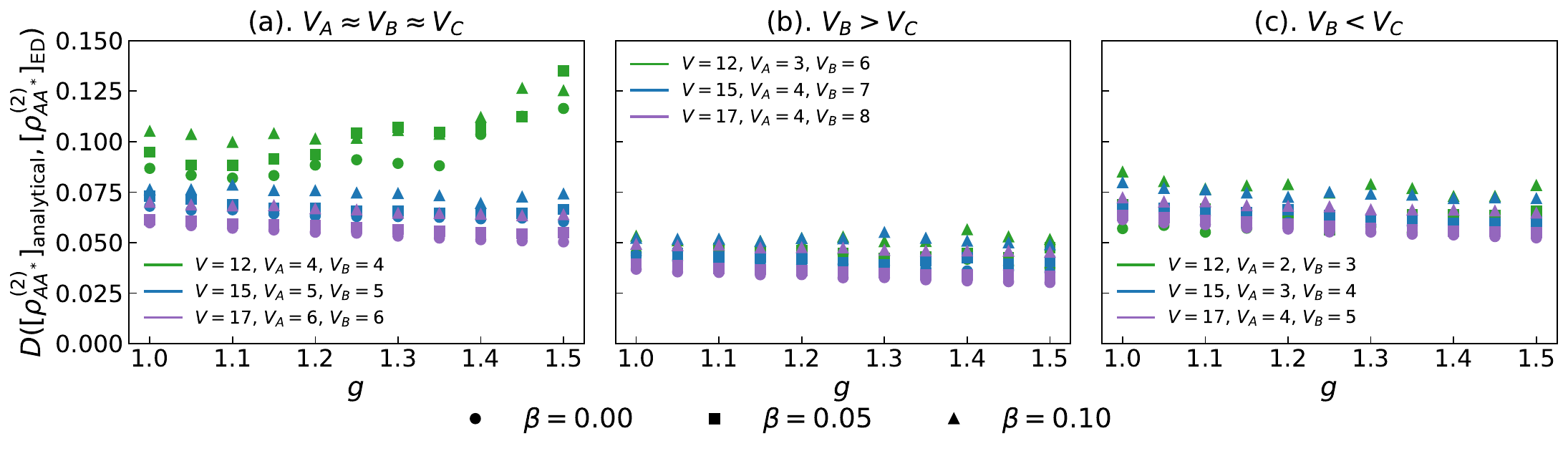}

    \caption{Trace distance between $\average{\rho_{AA^*}^{(2)}}_{\text{analytical}}$ and $\average{\rho_{AA^*}^{(2)}}_{\text{ED}}$ in the mixed-field Ising model with $h=0.35$ for three different tripartitions: (a) $V_A \approx V_B \approx V_C$, (b) $V_B \gg V_C$, and (c) $V_B \ll V_C$ 
    An average over $k = 500$ states is taken for $\average{\rho_{AA^*}^{(2)}}_{\text{ED}}$. Different inverse temperatures are represented by different symbols: 
    circles, squares, and triangles correspond to $\beta = 0$, $\beta = 0.05$, and $\beta = 0.1$, respectively. 
    }
    \label{fig:trace_distance_avg_theoretical}
\end{figure*}
\begin{figure}[htbp]
    \centering

    \includegraphics[width=0.5 \textwidth]{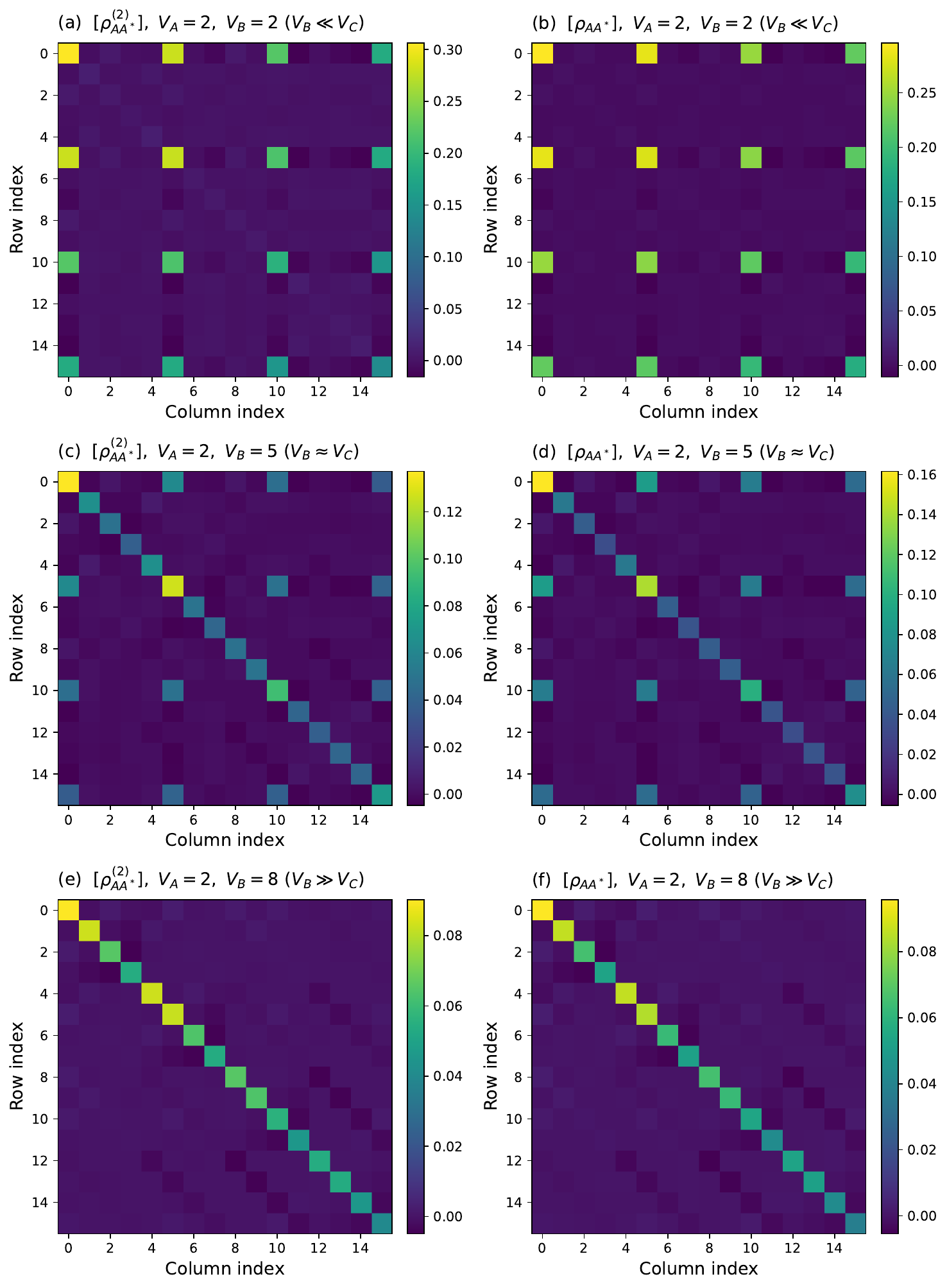}

    \caption{Visualization of matrix structures of (a,c,e) average $\rho_{AA^*}^{(2)}$ and (b,d,f) average $\rho_{AA^*}$ in the mixed-field Ising model with $(g,h)=(1.1,0.35)$ for different subsystem size regimes. The total system size is $V=12$. The average is taken over $k = 500$ eigenstates around inverse temperature $\beta = 0.1$. }
    
    \label{fig:rho_AA_visualization_all}
\end{figure}

Having provided evidence for the universality of the structure of $\rho_{AA^*}^{(2)}$, we now move to a test of our analytic predictions for said structure.
To examine how well the analytical expression  \eqref{eq:average_rho_AA_2_analytical} describes the average $\rho_{AA^*}^{(2)}$, we compare the analytical expression with the exact diagonalization results for $\rho_{AA^*}^{(2)}$ in the mixed-field Ising model 
of Eq.~\eqref{mixed_field_Ising_model}. 
Specifically, we compute the trace distance between the analytical prediction for the averaged density matrix, 
$\average{\rho_{AA^*}^{(2)}}_{\text{analytical}}$ 
\eqref{eq:average_rho_AA_2_analytical}
and the average obtained \textit{via} exact diagonalization $\average{\rho_{AA^*}^{(2)}}_{\text{ED}}$, denoted as $D(\average{\rho_{AA^*}^{(2)}}_{\text{analytical}}, \average{\rho_{AA^*}^{(2)}}_{\text{ED}})$. 
The ensemble average $[\cdots]_{\mathrm{ED}}$ is taken over the eigenstates whose energies 
fall in the window $[E-\Delta E, E+\Delta E]$ about the energy $E$ by fixed the 
target temperature $\beta$, the same microcanonical window used throughout this section.
We consider three types of tripartitions, corresponding to the three cases $V_B \gg V_C$, $V_C \gg V_B$ and $V_B \sim V_C$ respectively. 
We consider the case when $V_A \leq V_B$, and examine three system sizes $V = 12, 15, 17$.

We examine the following Hamiltonian parameters: the longitudinal field is set to be $h = 0.35$, while we consider values of the transverse field in the range $g \in [1,1.5]$.
We consider three inverse temperatures $\beta = 0, 0.05$ and $0.1$.
In this parameter and temperature range, the analytical expression of $S_2^{(3)}$ and the exact diagonalization average match for all three system sizes (see Fig.\ \ref{fig:bipartite_tripartite_compare}). 
Similar to when calculating $\average{S_2^{(3)}}_{\text{analytical}}$, the density of states used in the analytical expression $\average{\rho_{AA^*}^{(2)}}$ is obtained from exact diagonalization of the subsystem Hamiltonians. 

As shown in Fig.\ \ref{fig:trace_distance_avg_theoretical}, the analytical expression describes $\average{\rho_{AA^*}^{(2)}}$ reasonably well across all subsystem partitions and inverse temperatures examined, and the trace distance decreases with increasing system size.
Recall that, as discussed above, when $V_C > V_{AB}$, the standard subsystem ETH ansatz implies $\rho_{AB}$ takes a universal form \cite{Dymarsky_2018}, which suggests $\rho_{AA^*}^{(2)}$ should as well; the novel regime is the one in which $V_C < V_{AB}$, where subsystem ETH no longer applies. 
In order to focus on this regime, in Fig.~\ref{fig:trace_distance_avg_theoretical}, we therefore keep the subsystem ratios $V_A/V$ and $V_B/V$ non-decreasing as we increase system size, rather than having $V_C/V$ increase as system size increases.
The observed decrease in trace distance with system size provides evidence supporting that the analytical expression  \eqref{eq:average_rho_AA_2_analytical} approximates $\average{\rho_{AA^*}^{(2)}}$ well in the thermodynamic limit. 
Among the three subsystem partitions, finite-size deviations from the analytical expression are the most pronounced in the equal tripartition case $V_A \approx V_B \approx V_C$. 
This is expected because, in this regime, both $C_{\text{diag}}$ and $C_{\text{block}}$ contribute at the leading order, whereas only one of the two dominates in the other regimes.
Importantly, the agreement at the level of 
$\average{\rho_{AA^*}^{(2)}}$ provides a stronger check: the energy-constrained Haar random ensemble not only captures the average tripartite R\'enyi-2 multi-entropy (which we recall, is essentially also the $(2, 2)$-R\'enyi reflected entropy), but is also capable of describing the average reduced density matrix itself.

Finally, while we did not make any analytic predictions for the structure of the von Neumann limit of the reduced density matrix, $\rho_{AA^*}$, we nevertheless expect its structure to be qualitatively similar to that of $\rho_{AA^*}^{(2)}$.
Indeed, though the quantitative values of matrix entries in $\average{\rho_{AA^*}}$ is different from that in $\average{\rho_{AA^*}^{(2)}}$, the average of $\rho_{AA^*}$ is also dominated by diagonal terms when $L_B \gg L_C$, by ``block" terms when $L_B \ll L_C$, and both terms are present when $L_B = L_C$. 
This can be seen clearly from the side-by-side visual representation of the matrix structure for $\average{\rho_{AA^*}^{(2)}}$ and $\average{\rho_{AA^*}}$ in Fig.\ \ref{fig:rho_AA_visualization_all}.
This provides evidence for the picture discussed in Sec.~\ref{sec:tfd-tmd} that 
$\rho_{AA^*}$ itself indeed interpolates, at least at a qualitative level, between the 
product, thermo-mixed, and theromfield double states,  while 
$\rho_{AA^*}^{(2)}$ interpolates between closely related states.

\section{Summary and Discussion}
\label{summary_and_discussion}

In this work, we investigated the tripartite entanglement structure of chaotic eigenstates through two quantities that probe tripartite correlations: the R\'enyi multi-entropy $S_n^{(3)}$ and the R\'enyi reflected entropy $S_{m,n}^R$. 
Using an energy-constrained Haar random ansatz (Eq.\ \eqref{ergodic_tripartition}), we derived analytical leading-order expressions for the ensemble averages of these two quantities.
Beyond the scalar quantities, we also derived an analytical expression for the average of $\rho_{AA^*}^{(2)}$, which is relevant to the $(2, n)$-R\'enyi reflected entropy and operator entanglement, finding that it interpolates between three qualitatively distinct forms, related to the thermofield and thermo-mixed doubles.

Benchmarking these predictions against exact diagonalization of the mixed-field Ising model, we found that the energy-constrained Haar random ensemble describes not only the average R\'enyi-2 multi-entropy (which coincides up to a bipartite term with the (2, 2)-R\'enyi reflected entropy), but also the full average $\rho_{AA^*}^{(2)}$ at the level of trace distance.

Moreover, in the regime of nearly equal tripartition where correlations beyond bipartite or GHZ-type correlations are present, the range of Hamiltonian parameters over which the tripartite multi-entropy agrees with the constrained-Haar prediction closely tracks the corresponding range for the bipartite R\'enyi-2 entropy. We therefore find no evidence that genuinely tripartite correlations lead to an earlier breakdown of the constrained-Haar description than that observed for bipartite entanglement. This consistency in the range of validity does not imply identical finite-size behavior, and the deviations outside this regime depend quantitatively on the entanglement measure and subsystem partition.

We end the discussion with several open directions. 
First, recent work on bipartite entanglement has shown that higher moments of the von Neumann entropy provide a finer diagnostic for the degree of quantum chaos than the mean alone. 
Specifically, Ref.\ \cite{Rodriguez_Nieva_2024} studied the Kullback-Leibler divergence between the distribution of von Neumann entropies of the midspectrum eigenstates and that of an appropriately constrained Haar random ensemble. The Kullback-Leibler divergence measures the difference of the first and second moments of the entropy distribution in these two ensembles on the exponentially small scale set by the standard deviation of von Neumann entropy in the constrained random Haar ensemble.
In the mixed-field Ising model, the Kullback-Leibler divergence was found to be small only within a narrow maximally chaotic regime, and to increase exponentially when the Hamiltonian parameters are tuned away from it. 

A natural extension of the present work is to characterize the analogous Kullback-Leibler divergence for the distributions of (R\'enyi) multi-entropy and reflected entropy, and to determine whether the region of agreement with the constrained random Haar ensemble coincides with the maximally entangled region identified in Ref~\cite{Rodriguez_Nieva_2024}.
Since these tripartite quantities encode correlations not captured by bipartite entropies, such a comparison could either reinforce the status of this regime as a universal maximally chaotic window, or reveal a distinct and possibly narrower region of maximal chaos visible only to multipartite diagnostics. 

A second question concerns the finite-size corrections to the constrained Haar random prediction of $\average{S_2^{(3)}}$ and $\average{S_2}$. 
As observed in Section\ \ref{section: comparison_bipartite_tripartite}, although $\average{S_2^{(3)}}$ and $\average{S_2}$ exhibit a common near-unity plateau in the chaotic regime, their deviations away from this plateau do not collapse onto a single curve. 
This suggests that the finite-size corrections is partition-dependent and is not governed by a simple overall suppression of entanglement. 
It might be interesting to further explore the difference in the rate of convergence of the bipartite and tripartite entanglement quantities to the constrained Haar random prediction.

Finally, we comment on the prospects for experimentally probing the tripartite structures investigated here. 
The past decade has seen remarkable progress in the direct measurement of entanglement in many-body systems. 
The R\'enyi-2 entanglement entropy was accessed experimentally for systems of interacting delocalized particles using quantum interference of many-body twins in Ref. \cite{Islam_2015}. 
Shortly afterwards Ref. \cite{Kaufman_2016} applied similar methods to examine post-quench local thermalization in a system described by the Bose-Hubbard Hamiltonian, and observed an equivalence between the measured R\'enyi-2 entanglement entropy and the expected thermal entropy value consistent with ETH.

A complementary protocal via randomized measurements \cite{Brydges_2019, Elben_2019} was proposed and utilized to obtain R\'enyi-2 entanglement entropy in a trapped-ion quantum simulator.
The same experimental data can be post-processed to extract other quantities measuring entanglement:
protocols have been proposed to measure (R\'enyi) entanglement negativity \cite{Elben_2020} and (R\'enyi) operator entanglement \cite{Aniket_Rath_2023} and demonstrated using the experimental data in Ref. \cite{Brydges_2019}. 
The experimental measurement of operator entanglement is particularly relevant to this work, since the reduced density matrix of the R\'enyi canonical purification, $\rho_{AA^*}^{(2)}$, studied here is the object whose spectrum controls the operator entanglement of $\rho_{AB}$.
More generally, given the current development in entanglement measurement, it is plausible that the tripartite entanglement structures probed by R\'enyi reflected entropy and R\'enyi multi-entropy could come within experimental reach in the near future.

\section*{Acknowledgements}
S.R. is supported by a Simons
Investigator Grant from the Simons Foundation (Award
No. 566116). This work is supported by the Gordon
and Betty Moore Foundation EPiQS initiative, Grant
GBMF8685.01.
We acknowledge the use of generative AI in the production of Fig.~\ref{fig:phase-diagram}.

\appendix 

\section{Saddle Point Analysis for Multi-Entropy}

\subsection{Saddle point structure of multi-entropy}
\label{appendix:symmetric_saddle_multi_entropy}

In this appendix, we discuss the saddle point structure of $\average{Z_n^{(3)}}$ (Eq.\ \eqref{Z_n_3 average}), 
and show that under the physically reasonable assumption of strictly concave microcanonical entropy, the saddle point, if exists, is unique and has symmetric replica subsystem energies.
Since R\'enyi multi-entropy is symmetric among all three subsystems, we focus on the regime where $V_C \gg V_A, V_B$, and the other regimes follow the same argument. 

First, we write the microcanonical entropy of the subsystem $K$ ($K = A, B, C$) in terms of the microcanonical entropy density $s(u^K)$ and the energy density $u^K = E^K/V_K$:
\begin{equation}
    S_K(E^K) = V_K s(u^K).
    \label{microcanonical_entropy_as_entropy_density}
\end{equation}
Here, we assume the same form of microcanonical entropy density for all three subsystems, as well as the total system (i.e., for the total system, $S(E) = Vs(u)$ where $u = E/V$).

For clarity, we write down the expression for $\average{Z_n^{(3)}}$ in the regime where $V_C \gg V_A, V_B$ again here:
\begin{widetext}
\begin{equation}
    \average{Z_n^{(3)}} = \frac{1}{\mathcal{N}^{n^2}} 
       \sum_{E^A_1,\ldots, E^A_n}\sum_{E^B_1,\ldots, E^B_n} e^{\sum_{i=1}^nS_A(E^A_i) + \sum_{j=1}^nS_B(E^B_j) + \sum_{i,j=1}^nS_C(E - E^A_i - E^B_j)},
\end{equation}
\end{widetext}
where $\mathcal{N}$ is the normalization factor presented in Eq.\ \eqref{normalization constant}.

To perform the saddle point approximation for the sum in the numerator, we need to find the stationary point of the exponent. 
The exponent can be written in terms of $u^K$ and $s(u^K)$ as
\begin{equation}
    \Phi(\vec{u}) = \sum_{i=1}^n V_A s(u_i^A) + \sum_{j=1}^n V_B s(u_j^B) + \sum_{i, j=1}^n V_C s(u_{i,j}^C).
    \label{multi_entropy_exponent}
\end{equation}
Here, $\Phi(\vec{u})$ is a function of $\vec{u} = (u_1^A, u_2^A, \ldots, u_n^A, u_1^B, u_2^B, \ldots, u_n^B) \in \mathbb{R}^{2n}$, and $u_{i, j}^C$ can be expressed in terms of $u_i^A$ and $u_j^B$ as
\begin{equation}
    u_{i,j}^C = \frac{Vu - V_A u_i^A - V_B u_j^B}{V_C},
    \label{energy_conservation_subsystem_energy_density}
\end{equation}
which follows from the energy conservation equation $E^A + E^B + E^C = E$.

\subsubsection{Uniqueness of the stationary point}

First, we show that if the microcanonical entropy density is strictly concave (which is a reasonable assumption for chaotic systems away from phase transitions \footnote{Strict concavity of microcanonical entropy density implies positive heat capacity $C_V$.}), $\Phi(\vec{u})$ is strictly concave 
\footnote{Strict concavity is defined as follows: a real-valued function $f$ defined on a convex set $D$ is strictly concave if $f((1-\alpha)x + \alpha y) > (1-\alpha)f(x) + \alpha f(y)$ for all $x, y \in D$, $x \neq y$, and $\alpha \in (0, 1)$.}. It then follows that the stationary point of $\Phi(\vec{u})$, if exists, is a maximizer and is unique.

That $\Phi(\vec{u})$ is strictly concave follows directly from the strict concavity of the microcanonical entropy densities. 
Specifically, consider $\vec{u} = (u_1^A, u_2^A, \ldots, u_n^A, u_1^B, u_2^B, \ldots, u_n^B)$, $\vec{v} = (v_1^A, v_2^A, \ldots, v_n^A, v_1^B, v_2^B, \ldots, v_n^B)$, and $\vec{w} = \alpha \vec{u} + (1-\alpha) \vec{v}$, where $\alpha \in (0, 1)$. 
We write $w_{i}^A = \alpha u_i^A + (1-\alpha)v_i^A$, and $w_{j}^B = \alpha u_j^B + (1-\alpha)v_j^B$ ($i, j = 1, \ldots, n$).
By the energy conservation requirement in Eq.\ \eqref{energy_conservation_subsystem_energy_density}, 
\begin{equation}
    w_{i, j}^C = \frac{V u - V_A u_i^A - V_B u_j^B}{V_C} = \alpha u_{i, j}^C + (1-\alpha) v_{i, j}^C,
\end{equation}
where $u = E/V$ is the energy density in the total system.

Thus, assuming strict concavity of $s(u)$, if $u_i^A \neq v_i^A$, 
\begin{equation}
    s(w_i^A) > \alpha s(u_i^A) + (1-\alpha) s(v_i^A),
\end{equation}
and similarly $s(w_j^B) > \alpha s(u_j^B) + (1-\alpha) s(v_j^B)$ if $u_j^B \neq v_j^B$, and $s(w_{i, j}^C) > \alpha s(u_{i, j}^C) + (1-\alpha) s(v_{i, j}^C)$ if $u_{i, j}^C \neq v_{i, j}^C$. From the form of $\Phi(\vec{u})$ in Eq.\ \eqref{multi_entropy_exponent}, it then follows directly that if $\vec{u} \neq \vec{v}$, 
\begin{equation}
    \Phi(\alpha \vec{u} + (1-\alpha)\vec{v}) > \alpha \Phi(\vec{u}) + (1-\alpha)\Phi(\vec{v}).
\end{equation}

Since $\Phi(\vec{u})$ is a concave function, any stationary point must be a maximizer. 
A strictly concave function defined on a convex set (like $\mathbb{R}^{2n}$ or a Cartesian product of intervals in $\mathbb{R}$) has at most one maximizer. 
Thus, the stationary point of $\Phi(\vec{u})$, if exists, is a maximizer and is unique.

\subsubsection{Replica symmetry of the saddle point}

To show the replica symmetry of the saddle point, we first note that the exponent $\Phi(\vec{u})$ (Eq.\ \eqref{multi_entropy_exponent}) is invariant under any permutation of $(u_1^A, \ldots, u_n^A)$ and $(u_1^B, \ldots, u_n^B)$. 
Specifically, we consider the group of cyclic permutation $G = C_n \times C_n$ generated by $\tau_A \times e_B$ and $e_A \times \tau_B$, where
\begin{align}
    \nonumber 
    &\tau_A: (u_1^A, u_2^A, \ldots, u_n^A) \to (u_2^A, u_3^A, \ldots, u_1^A), \\
    &\tau_B: (u_1^B, u_2^B, \ldots, u_n^B) \to (u_2^B, u_3^B, \ldots, u_1^B).
\end{align}
$\Phi(\vec{u})$ is invariant under any transformation $g \in G$, i.e., $\Phi(g(\vec{u})) = \Phi(\vec{u})$.

Consider the average $\vec{u}_{\text{avg}}$ of all transformations $g(\vec{u})$ on a certain $\vec{u} = (u_1^A, u_2^A, \ldots, u_n^A, u_1^B, u_2^B, \ldots, u_n^B)$:
\begin{equation}
    \vec{u}_{\text{avg}} = \frac{1}{|G|} \sum_{g \in G} g(\vec{u}),
\end{equation}
where $|G|$ is the cardinality of the group of cyclic transformations $C_n \times C_n$. It is easy to see from the form of transformations in $G$ that 
\begin{equation}
    u_{i, \text{avg}}^A = \frac{1}{n}\sum_{i=1}^n u_i^A, \quad  u_{j, \text{avg}}^B = \frac{1}{n} \sum_{j=1}^n u_j^B.  
    \label{form_of_u_avg}
\end{equation}

Since $\Phi(\vec{u})$ is concave, 
\begin{equation}
    \Phi(\vec{u}_{\text{avg}}) = \Phi(\frac{1}{|G|} \sum_{g \in G} g(\vec{u})) \geq \frac{1}{|G|} \sum_{g \in G} \Phi(g(\vec{u})).
\end{equation}
Moreover, since $\Phi(\vec{u})$ is invariant under any transformation $g \in G$, 
\begin{equation}
    \Phi(\vec{u}_{\text{avg}}) \geq \frac{1}{|G|} \sum_{g \in G} \Phi(g(\vec{u})) = \Phi(\vec{u}).
    \label{equality of average exponent}
\end{equation}

Now, assume that there exists $\vec{u}$ that maximizes the exponent $\Phi(\vec{u})$. By Eq.\ \eqref{equality of average exponent}, $\Phi(\vec{u}_{\text{avg}}) \geq \Phi(\vec{u})$. Since $\vec{u}$ is a maximizer of $\Phi$, the equality must hold, and $\vec{u}_{\text{avg}}$ is also a maximizer of $\Phi(\vec{u})$. From Eq.\ \eqref{form_of_u_avg}, it is clear that $\vec{u}_{\text{avg}}$ exhibits symmetry among subsystem replica energies, i.e., $u_i^A = u^A$ for all $i = 1, \ldots, n$ and $u_j^B = u^B$ for all $j = 1, \ldots, n$.

Combining with the discussions in the previous subsection, we conclude that the maximizer of the exponent in the expression of R\'enyi multi-entropy, if exists, is unique and replica-symmetric.

\subsection{Saddle-point approximation and curvature}
\label{appendix:multi_entropy_curvature}

In this appendix, following the same method as was used in Ref.\ \cite{Lu_2017}, we examine the curvature of R\'enyi multi-entropy using the saddle point approximation. Since $\average{Z_n^{(3)}}$ and $\average{Z_{n, \text{RS}}^{(3)}}$ share the same saddle point, for simplicity, we derive the saddle point equations from the expressions for $\average{Z_{n, \text{RS}}^{(3)}}$ (Eq.\ \eqref{Z_n_3_average_n_approx}). We focus on the regime where $V_C \gg V_A, V_B$, and the other regimes should exhibit the same behavior by the symmetry of multi-entropy among subsystems.

As a baseline for comparison, we summarize the behavior of $\average{S_n^{(3)}}$ in Haar random states. 
The leading order of $\average{S_n^{(3)}}$ takes the form \cite{Iizuka:2024pzm}
\begin{align}
    \average{S_n^{(3)}} = 
    \begin{cases}
        V_A + V_B, \quad V_C \gg V_A, V_B \\
        V_A + V_C, \quad V_B \gg V_A, V_C \\
        V_B + V_C, \quad V_A \gg V_B, V_C
    \end{cases}
    \label{S_n_3_random_pure_states}
\end{align}
The ``much greater than" sign becomes a ``greater than" sign when $n=2, 3$. 
For Haar random states, as shown in the expressions above, when $V_C \gg V_A, V_B$, $\average{S_n^{(3)}}$ depends linearly on the size of the combined subsystem $\mathcal{H}_A \otimes \mathcal{H}_B$, and is independent of the internal partition of this combined system into $\mathcal{H}_A$ and $\mathcal{H}_B$. 
In other words, $\average{S_n^{(3)}}$ depends linearly on $f = f_A + f_B$, and is independent of $f_A - f_B$ when $f_A + f_B$ is held fixed (the subsystem fraction $f_K = V_K/V$, $K = A, B$). 

Now we consider R\'enyi multi-entropy in states described by Eq.\ \eqref{ergodic_tripartition}. When written in terms of microcanonical entropy densities (Eq.\ \eqref{microcanonical_entropy_as_entropy_density}) and subsystem fractions, in the regime where $V_C \gg V_A, V_B$, 
\begin{widetext}
\begin{equation}
    \average{Z_n^{(3)}} = \frac{\sum_{u^A}\sum_{u^B} e^{nf_AVs(u^A) + nf_BVs(u^B)+n^2(1-f_A-f_B)Vs(u^C)}}{(\sum_{u^A}\sum_{u^B}e^{f_AVs(u^A) + f_BVs(u^B) + (1-f_A-f_B)Vs(u^C)})^{n^2}},
\end{equation}
\end{widetext}
where the energy densities satisfy
\begin{equation}
    f_Au^A + f_Bu^B + (1-f_A-f_B)u^C = u.
    \label{u_energy_conservation_constraint}
\end{equation}

Applying the saddle point approximation, 
$\average{S_n^{(3)}} = \frac{1}{1-n}\frac{1}{n} \ln \average{Z_n^{(3)}}$ takes the form:
\begin{align}
    \average{S_n^{(3)}} & = \frac{V}{1-n}
    \big\{f_A s(u^{A*}) + f_B s(u^{B*}) \nonumber \\ &
    \quad 
    + n(1-f_A-f_B)s(u^{C*})-ns(u)\big\},
    \label{saddle_point_equation_Sn}
\end{align}
where $u^{A*}, u^{B*}$ and $u^{C*}$ satisfy the saddle point equations:
\begin{align}
    & \frac{\partial s(u)}{\partial u}\vert_{u=u^{A*}} = n\frac{\partial s(u)}{\partial u}\vert_{u = u^{C*}},     \label{u_saddle_point_constraint} \nonumber
    \\ 
    & \frac{\partial s(u)}{\partial u}\vert_{u=u^{B*}} = n\frac{\partial s(u)}{\partial u}\vert_{u = u^{C*}}.
\end{align}

Due to the concavity of entropy densities, $\partial s(u)/\partial u$ is monotonic and thus injective. 
Thus, we can set
\begin{equation}
    u^{A*} = u^{B*} = u_{AB}^*.
\end{equation}
Denoting the subsystem fraction of $\mathcal{H}_A \otimes \mathcal{H}_B$ as $f$:
\begin{equation}
    f_A + f_B = f,
\end{equation}
the average of R\'enyi multi-entropy becomes
\begin{equation}
    \average{S_n^{(3)}} = \frac{V}{1-n}\big\{fs(u_{AB}^*) + n(1-f)s(u^{C*}) - ns(u)
    \big\},
    \label{S_n_3_two_system}
\end{equation}
with saddle point equation and energy constraint:
\begin{align}
    & \frac{\partial s(u)}{\partial u}\vert_{u = u_{AB}^*} = n\frac{\partial s(u)}{\partial u}\vert_{u = u^{C*}}, \nonumber 
    \\
    & fu_{AB}^* + (1-f)u^{C*} = u.
    \label{S_n_3_2_system_constraints}
\end{align}

The form of $\average{S_n^{(3)}}$ (Eq.\ \eqref{S_n_3_two_system}) shows that in this regime, R\'enyi multi-entropy is insensitive to the internal partition of $\mathcal{H}_A \otimes \mathcal{H}_B$ in chaotic eigenstates.
This is consistent with the case of Haar random states (Eq.\ \eqref{S_n_3_random_pure_states}).

To check the curvature of $\average{S_n^{(3)}}$ in the $f = f_A + f_B$ direction, we compute $\partial^2 \average{S_n^{(3)}}/\partial f^2$. 
The first derivative of $\average{S_n^{(3)}}$ with respect to $f$ takes the form
\begin{widetext}
  \begin{align}
    \frac{\partial \average{S_n^{(3)}}}{\partial f} & = \frac{V}{1-n}\Big\{s(u_{AB}^*) - ns(u^{C*}) + f\frac{\partial s(u_{AB}^*)}{\partial u_{AB}^*}\frac{\partial u_{AB}^*}{\partial f} + n(1-f)\frac{\partial s(u^{C*})}{\partial u^{C*}}\frac{\partial u^{C*}}{\partial f}
    \Big\} \\ \nonumber
    & =  \frac{V}{1-n}\Big\{s(u_{AB}^*) - ns(u^{C*}) + \frac{\partial s(u_{AB}^*)}{\partial u_{AB}^*}(f\frac{\partial u_{AB}^*}{\partial f} + (1-f)\frac{\partial u^{C*}}{\partial f})
    \Big\} \\ \nonumber
    & = \frac{V}{1-n}
    \Big\{s(u_{AB}^*) - ns(u^{C*}) + \frac{\partial s(u_{AB}^*)}{\partial u_{AB}^*}(u^{C*} - u_{AB}^*)
    \Big\}.
\end{align}  
\end{widetext}
Here, the second equality holds due to the saddle point equation, and the third equality holds due to the equality
\begin{equation}
    f\frac{\partial u_{AB}^*}{\partial f} + (1-f)\frac{\partial u^{C*}}{\partial f} = u^{C*} - u_{AB}^*,
    \label{derivative_of_energy_constraint}
\end{equation}
which is obtained by taking the derivatives of both sides of the energy constraint (Eq.\ \eqref{S_n_3_2_system_constraints}) with respect to $f$.
Taking the derivative with respect to $f$ again and simplify using the saddle point equation, the expression becomes
\begin{equation}
    \frac{\partial^2 \average{S_n^{(3)}}}{\partial f^2} = \frac{V}{1-n}\frac{\partial^2 s(u_{AB}^*)}{\partial (u_{AB}^*)^2}\frac{\partial u_{AB}^*}{\partial f}(u^{C*} - u_{AB}^*).
\end{equation}
Since $s(u_{AB}^*)$ is concave, $\frac{\partial^2 s(u_{AB}^*)}{\partial (u_{AB}^*)^2} < 0$. 
Taking derivative on both sides of the saddle point equation with respect to $f$, it follows that $\frac{\partial u_{AB}^*}{\partial f}$ and $\frac{\partial u^{C*}}{\partial f}$ have the same sign. 
Combined with Eq.\ \eqref{derivative_of_energy_constraint}, we can see that $\frac{\partial u_{AB}^*}{\partial f}$ and $u^{C*} - u_{AB}^*$ have the same sign. Since we are considering the regime where $n > 1$, 
\begin{equation}
    \frac{\partial^2 \average{S_n^{(3)}}}{\partial f^2} > 0.
\end{equation}
This demonstrates that consistent with the behavior of R\'enyi entropy, R\'enyi multi-entropy is convex along the $f = f_A + f_B$ direction when the R\'enyi index $n > 1$. Such convexity was not observed for $\average{S_n^{(3)}}$ in Haar random states.

\section{$\average{Z_{n}^{(3)}}$ and  $\average{Z_{n, \text{RS}}^{(3)}}$ for Gaussian Density of States}

\subsection{Difference between $\average{Z_{n}^{(3)}}$ and  $\average{Z_{n, \text{RS}}^{(3)}}$}
\label{appendix:multi_entropy_exact_RS_comparison}

In this appendix, we compute $\average{Z_{n}^{(3)}}$ and $\average{Z_{n, \text{RS}}^{(3)}}$ for systems with Gaussian density of states and compare their differences.
Specifically, we consider microcanonical entropy of the following form
\begin{equation}
    S_K(E^K) = V_K \ln 2 - \frac{1}{2}\frac{(E^K)^2}{V_K} - \frac{1}{2}\ln (2\pi V_K).
    \label{normalized_S_K}
\end{equation}
Here, $K$ labels the subsystems $A, B$ and $C$. We assume that the entropy density in each subsystem, as well as the total system, is the same.
The last term in Eq.\ \eqref{normalized_S_K} is a normalization that ensures that the Hilbert space dimension of a subsystem with $V_K$ number of sites is $2^{V_K}$: $\int dE^K e^{S_K(E^K)} = 2^{V_K}$.

We focus on the case with $V_C \gg V_A, V_B$. The other two cases follow directly due to symmetry among the three subsystems. 
In the continuous limit, 
\begin{widetext}
    \begin{align}
    \average{Z_{n, \text{RS}}^{(3)}} &=  \frac{1}{\mathcal{N}^{n^2}} \int_{-\infty}^\infty dE^A \int_{-\infty}^\infty dE^B e^{nS_A(E^A) + nS_B(E^B) + n^2 S_C(E^C)} 
    \label{Z_n_3_approx_Gaussian}
    \\ \nonumber
    &=  \frac{1}{\mathcal{N}^{n^2}} 
    \frac{ 2^{nV_A + nV_B + n^2V_C} }{((2\pi)^2 V_A V_B)^{n/2}(2\pi V_C)^{n^2/2}} 
    \frac{2\pi}{n} e^{-\frac{n^2E^2}{2(nV_A + nV_B + V_C)}} \sqrt{\frac{V_AV_BV_C}{nV_A+nV_B+V_C}},
\end{align}
where $\mathcal{N}$ is the normalization factor (Eq.\ \eqref{normalization constant}) and can be evaluated as 
\begin{equation}
    \mathcal{N} = \int dE^A \int dE^B e^{S_A(E^A) + S_B(E^B) + S_C(E - E^A - E^B)} = \frac{1}{\sqrt{2\pi V}}2^Ve^{-E^2/(2V)}.
    \label{normalization_factor_Gaussian}
\end{equation}

On the other hand, 
\begin{align}
     \average{Z_{n}^{(3)}} &= \frac{1}{\mathcal{N}^{n^2}} \prod_{i=1}^n \int_{-\infty}^\infty dE^A_i \prod_{j=1}^n \int_{-\infty}^\infty dE_j^B e^{\sum_{i=1}^n S_A(E^A_i) + \sum_{j=1}^n S_B(E^B_j) + \sum_{i,j=1}^n S_C(E - E^A_i - E^B_j)} \\ \nonumber 
    & =  \frac{1}{\mathcal{N}^{n^2}} 
    \frac{ 2^{nV_A + nV_B + n^2V_C} }{((2\pi)^2 V_A V_B)^{n/2}(2\pi V_C)^{n^2/2}} 
    \prod_{i=1}^n \int_{-\infty}^\infty dE^A_i \prod_{j=1}^n \int_{-\infty}^\infty dE^B_j e^{-\frac{1}{2}(\sum_{i=1}^n \frac{(E^A_i)^2}{V_A} + \sum_{j=1}^n\frac{(E^B_j)^2}{V_B} + \sum_{i,j=1}^n \frac{(E - E^A_i - E^B_j)^2}{V_C})}.
\end{align}
To evaluate this $2n$-dimensional integral, we first perform a change of variables.
Consider the averages $E^A = \frac{1}{n}\sum_{i=1}^n E^A_i$ and $E^B = \frac{1}{n}\sum_{j=1}^n E^B_j$, and the differences $a_i = E^A_i - E^A$ and $b_j = E^B_j - E^B$ which satisfy $\sum_{i=1}^n a_i = 0$ and $\sum_{j=1}^n b_j = 0$.
The integral can be rewritten as 
    \begin{align}
    \average{Z_{n}^{(3)}} &=  \frac{1}{\mathcal{N}^{n^2}} 
    \frac{ 2^{nV_A + nV_B + n^2V_C} }{((2\pi)^2 V_A V_B)^{n/2}(2\pi V_C)^{n^2/2}} 
    (\int_{-\infty}^\infty dE^A \int_{-\infty}^\infty dE^B e^{-\frac{1}{2}(n\frac{(E^A)^2}{V_A} + n\frac{(E^B)^2}{V_B} + n^2\frac{(E - E^A - E^B)^2}{V_C})}) F \\ \nonumber
    &=  \average{Z_{n, \text{RS}}^{(3)}} \ F,
\end{align}
where $F$ is the product of two $(n-1)$-dimensional Gaussian integrals in the constrained subspace spanned by $a_i$ and $b_j$:
\begin{align}
    F &= n
    I_a I_b,
    \quad
    I_a:=
    \int_{\sum_i a_i = 0} \prod_{i=1}^n da_i e^{-\frac{1}{2} (\frac{1}{V_A} + \frac{n}{V_C}) \sum_{i=1}^n a_i^2},
    \quad
    I_b:=\int_{\sum_j b_j = 0} \prod_{j=1}^n db_j e^{-\frac{1}{2}(\frac{1}{V_B} + \frac{n}{V_C}) \sum_{j=1}^n b_j^2}. 
\end{align}
\end{widetext}
Here, the additional factor of $n$ comes from the change of variables.
If we re-express $(a_1,\ldots,a_n)$ as a vector living in the $(n-1)$-dimensional subspace constrained by $\sum_i a_i = 0$, the integral in $a_i$ is just the integral of the vector norm in $(n-1)$-dimensions. 
The Gaussian integrals can be straightforwardly evaluated as 
\begin{align}
     & 
    I_a
    = \frac{(2\pi)^{(n-1)/2}}{(\frac{1}{V_A} + \frac{n}{V_C})^{(n-1)/2}}, 
    \quad
     I_b
     = \frac{(2\pi)^{(n-1)/2}}{(\frac{1}{V_B} + \frac{n}{V_C})^{(n-1)/2}}.
\end{align}

Thus, the ratio between $\average{Z_{n}^{(3)}}$ and $\average{Z_{n, \text{RS}}^{(3)}}$ takes the form
\begin{align}
   \nonumber F & = \frac{\average{Z_{n}^{(3)}}}{\average{Z_{n, \text{RS}}^{(3)}}}  \\ 
   &= n(2\pi)^{n-1} \frac{(V_AV_B)^{(n-1)/2}V_C^{n-1}}{(nV_A + V_C)^{(n-1)/2}(nV_B + V_C)^{(n-1)/2}}.
    \label{Z_n_3_exact_approx_proportion}
\end{align}
In the limit $V \to \infty$, if $V_A/V$, $V_B/V$ and $V_C/V$ remain finite, the ratio scales as $V^{n-1}$: 
\begin{equation}
    F \propto V^{n-1}.
\end{equation}
Thus, from the definition of $S_n^{(3)}$, we see that
\begin{equation}
    \average{S_{n}^{(3)}} - \average{S_{n, \text{RS}}^{(3)}} \sim -\frac{1}{n}\ln V.
\end{equation}
Specifically, $\average{S_{n}^{(3)}}$ and $\average{S_{n, \text{RS}}^{(3)}}$ share the same behavior in the leading order $O(V)$, which is expected since these two expressions share the same saddle point, and leading-order behavior is completely determined by the saddle point. On the other hand, $\average{S_{n}^{(3)}}$ is smaller than $\average{S_{n, \text{RS}}^{(3)}}$ by a subleading correction of the order $O(\ln V)$. As shown in the computation above, this subleading correction arises from the fluctuations in the replica subsystem energy $E^K_i - \frac{1}{n}\sum_{i=1}^n E^K_i$.

\subsection{Comparison with Haar random states}
\label{appendix:multi_Haar_random_comparison}

To further investigate the meaning of the $O(\ln V)$ correction, we compare the average R\'enyi multi-entropy in states drawn from a Haar random ensemble subject to energy conservation (Eq.\ \eqref{ergodic_tripartition}) with that drawn from Haar random ensemble with no symmetry constraints. 
Our comparison focuses on the dominant Wick contraction. For R\'enyi entropies with integer $n \geq 2$, unlike the von-Neumann limit, the contribution from summing over non-leading Wick contractions is vanishing in the thermodynamic limit away from the phase transition point, and is $O(1)$ at the phase transition \cite{Dong_2020}. 
Since we are examining an $O(\ln V)$ difference which already appears when considering the leading Wick contraction, it is reasonable to ignore all other Wick contractions for this purpose. 

\subsubsection{R\'enyi bipartite entropy}

To get a sense of the comparison between the behaviors of R\'enyi entropy in energy-constrained Haar random states and that in  Haar random states in the thermodynamic limit, we first consider the bipartite case. 
To the leading-order, the R\'enyi-$n$ bipartite entropy takes the form \cite{Lu_2017}:
\begin{equation}
    \average{S_n} = \frac{1}{1-n} \ln \Bigg\{
        \frac{\sum_{E^A} e^{S_A(E^A) + nS_B(E - E^A)}}{(\sum_{E^A} e^{S_A(E^A) + S_B(E - E^A)})^n}
        \Bigg\}.
\end{equation}
In the continuous limit, the sum becomes an integral over $E^A$, and the evaluation is straightforward:
\begin{widetext}
    \begin{align}
    &
    \int_{-\infty}^\infty dE^A e^{S_A(E^A) + nS_B(E - E^A)} = \frac{1}{\sqrt{2\pi V_A}}\frac{1}{(\sqrt{2\pi V_B})^n} \sqrt{2\pi} \sqrt{\frac{V_AV_B}{V_B+nV_A}} 2^{V_A + nV_B} e^{-nE^2/2(nV_A + V_B)},
    \nonumber \\
    &
    \int dE^A e^{S_A(E^A) + S_B(E - E^A)} = \frac{1}{\sqrt{2\pi V}} 2^V e^{-E^2/2V}.
\end{align}

Substituting this into the expression for $S_n$, we find that
\begin{align}
    \nonumber \average{S_n} &=  V_A - \frac{nE^2 V_A}{2V(nV_A + V_B)} 
    + \frac{1}{1-n}\ln \Bigg\{
       \frac{1}{f_B^{(n-1)/2} (f_B + nf_A)^{1/2}}\Bigg\}.
\end{align}
We compare $\average{S_n}$ to the leading-order R\'enyi-$n$ entropy in Haar random states, which takes the form $S_{n, \text{Haar}} = V_A$. For the mid-spectrum eigenstates, the correction is $O(1)$, and $S_n < S_{n, \text{Haar}}$ (note that $f_A, f_B < 1$) which is physically reasonable. 
The $O(1)$ term here is not constant, but rather dependent on subsystem partitions.

\subsubsection{R\'enyi multi-entropy}

To compare the behaviors of R\'enyi multi-entropy in energy-constrained Haar random states and that in Haar random states, we first recall the expression for R\'enyi multi-entropy in Haar random states \cite{Iizuka:2024pzm}. 
The average multi-entropy $\average{Z_{n, \text{Haar}}^{(3)}}$ is expressed in a cycle-counting form: the Wick contraction corresponding to the permutation $\tau$ gives a contribution of $Z_n^{(3)} = (L_A^{\sigma(\tau^{-1} \circ g_A^{(n)})}L_B^{\sigma(\tau^{-1} \circ g_B^{(n)})}L_C^{\sigma(\tau^{-1})})/(L_AL_BL_C)^{n^2}$. 
Here, $L_K$ is Hilbert space dimension of the subsystem $K$, which we take to be $2^{V_K}$. Specifically, when $V_C \gg V_A, V_B$, $\average{S_{n, \text{Haar}}^{(3)}} = V_A + V_B$.

The replica-symmetric expression (Eq.\ \eqref{Z_n_3_approx_Gaussian}) gives the following form of average R\'enyi multi-entropy in energy-constrained Haar random states:
    \begin{equation}
    \average{S_{n, \text{RS}}^{(3)}} = (V_A + V_B) - \frac{nE^2(V_A + V_B)}{2V(nV_A + nV_B + V_C)} 
    + \frac{1}{n(1-n)}\ln \Bigg\{
        \frac{1}{n(2\pi)^{n-1}} 
        \frac{V^{n^2/2}}{(V_AV_B)^{(n-1)/2}V_C^{(n^2-1)/2}(nV_A + nV_B + V_C)^{1/2}}
        \Bigg\}.
\end{equation}
The second term is vanishing if $E$ scales slower than $\sqrt{V}$. 
However, by a simple power-counting, we find that the last term in $S_{n, \text{RS}}^{(3)}$ is of the order $\frac{1}{n(1-n)}\ln(V^{(1-n)})$, which corresponds to a $\frac{1}{n}\ln V$ enhancement to the R\'enyi multi-entropy as compared to the Haar random value. 

The $O(\ln V)$ negative departure of $\average{Z_n^{(3)}}$ from $\average{Z_{n, \text{RS}}^{(3)}}$ exactly cancels this spurious $O(\ln V)$ enhancement.
Specifically, we have shown that $\average{Z_n^{(3)}} = F \average{Z_{n, \text{RS}}^{(3)}}$, where $F$ is the contribution from the fluctuations in replica energies around their averages. 
Substituting the expression for $F$ (Eq.\ \eqref{Z_n_3_exact_approx_proportion}) into the $\average{S_{n, \text{RS}}^{(3)}}$ obtained above, we see that
\begin{equation}
\label{appendix_multi_entropy_O1}
    \average{S_{n}^{(3)}} = (V_A + V_B) 
    - \frac{nE^2(V_A + V_B)}{2V(nV_A + nV_B + V_C)} 
    + \frac{1}{n(1-n)} 
    \ln\Bigg\{
        \frac{f_C^{n-1}}{f_C^{(n^2-1)/2}(nf_A + nf_B + f_C)^{1/2}((nf_A + f_C)(nf_B+f_C))^{(n-1)/2}}
        \Bigg\}.
\end{equation}
\end{widetext}
This expression is different from the Haar random value $V_A + V_B$ only in $O(1)$ and higher-order, and a preliminary check shows that $\average{S_{n}^{(3)}} < S_{n, \text{Haar}}^{(3)}$.

\section{Saddle Point Analysis for R\'enyi Reflected Entropy}
\label{appendix:analytical_calculation_saddle_point_reflected_entropy}

\subsection{$V_A + V_B < V_C$}
\label{appendix:saddle_point_first_case_reflected_entropy}

First, we examine saddle point approximation for R\'enyi reflected entropy in the case when $V_A + V_B < V_C$. 
We write $S_K(E^K)$ as a function of $s(u_K)$ according to Eq.\ \eqref{microcanonical_entropy_as_entropy_density}. Then 
\begin{equation}
    \average{Z_{m, n}^R} = \frac{\sum_{u_A}\sum_{u_B} e^{nf_AVs(u_A) + nf_BVs(u_B) + mn(1-f_A-f_B)Vs(u_C)}}{(\sum_{u_A}\sum_{u_B}e^{f_AVs(u_A)+f_BVs(u_B)+(1-f_A-f_B)s(u_C)})^{mn}},
\end{equation}
where the energy densities satisfy the energy conservation constraint
\begin{equation}
    f_Au_A + f_Bu_B + (1-f_A-f_B)u_C = u.
    \label{energy_constraint_reflected}
\end{equation}
Applying the saddle point approximation, we obtain
\begin{align}
    \nonumber \frac{1}{1-n}\ln(\average{Z_{m, n}^R}) & = \frac{nV}{1-n}
    \Big\{f_As(u_A^*) + f_Bs(u_B^*) \\ & + m(1-f_A-f_B)s(u_C^*) - ms(u)\Big\},
    \label{saddle_point_Z_m_n_case_1}
\end{align}
where $u_A^*$, $u_B^*$ and $u_C^*$ satisfy the saddle point equations:
\begin{align}
    \nonumber \frac{\partial s(u)}{\partial u}\vert_{u = u_A^*} = m\frac{\partial s(u)}{\partial u}\vert_{u = u_C^*}, \\  \frac{\partial s(u)}{\partial u}\vert_{u = u_B^*} = m\frac{\partial s(u)}{\partial u}\vert_{u = u_C^*}.
\end{align}
Due to the concavity of $s(u)$, $\frac{\partial s(u)}{\partial u}$ is monotonic, leading to $u_A^* = u_B^*$, and we denote the energy densities as $u_{AB}^*$. 
Thus, Eq.\ \eqref{saddle_point_Z_m_n_case_1} can be simplified into
\begin{align}
    \nonumber \frac{1}{1-n}\ln(\average{Z_{m, n}^R}) = & \frac{nV}{1-n}
    \Big\{f_{AB}s(u_{AB}^*) \\ + &m(1-f_{AB})s(u_C^*) - ms(u)
    \Big\},
    \label{saddle_point_Z_m_n_case_1_simplified}
\end{align}
where $f_{AB} = f_A + f_B = (V_A + V_B)/V$. 
The corresponding energy constraint and saddle point equation also simplify into
\begin{align}
    & f_{AB}u_{AB}^* + (1 - f_{AB})u_C^* = u, \\
    & \frac{\partial s(u)}{\partial u}\vert_{u = u_{AB}^*} = m \frac{\partial s(u)}{\partial u}\vert_{u = u_C^*}.
\end{align}

On the other hand, $\average{Z_m}$ takes the form
\begin{equation}
    \average{Z_m} = \frac{\sum_{u_{AB}}e^{f_{AB}Vs(u_{AB})+m(1-f_{AB})Vs(u_c)}}{(\sum_{u_{AB}}e^{f_{AB}Vs(u_{AB})+(1-f_{AB})s(u_C)})^m},
\end{equation}
where the energy densities satisfy the energy conservation constraint
\begin{equation}
    f_{AB}u_{AB} + (1-f_{AB})u_C = u.
    \label{energy_constraint_bipartite}
\end{equation}
The saddle point approximation leads to 
\begin{align}
    \nonumber \frac{n}{1-n} \ln(\average{Z_m}) &=  \frac{nV}{1-n}
    \Big\{f_{AB}s(u_{AB}^*) \\ 
    &\quad
    +  m(1-f_{AB})s(u_C^*) - ms(u)
    \Big\},
\end{align}
with saddle point equations:
\begin{equation}
    \frac{\partial s(u)}{\partial u}\vert_{u = u_{AB}^*} = m\frac{\partial s(u)}{\partial u}\vert_{u = u_C^*}.
\end{equation}

Comparing the saddle point approximation and the corresponding constraints for $\average{Z_m}$ with that for $\average{Z_{m, n}^R}$ in Eq.\ \eqref{saddle_point_Z_m_n_case_1_simplified}, we can see that 
\begin{equation}
    \average{S_{m, n}^R(A\!:\!B)} = \frac{1}{1-n}(\ln (\average{Z_{m,n}^R}) - n\ln(\average{Z_m})) = 0.
\end{equation}
Thus, after the saddle point approximation, the leading-order R\'enyi reflected entropy vanishes when the size of subsystem $C$ (the bath) is larger than half of the total system size. 
This result is consistent with the behavior of (R\'enyi) reflected entropy in Haar random states.

\section{Supplement to Exact Diagonalization Results}

\subsection{Full expression for $\average{Z_{2, \text{exact}}^{(3)}}$}
\label{appendix:full_expression_Z_2_3}

When calculating $\average{Z_2^{(3)}}$ for comparison with exact diagonalization results, we use the full expression of $\average{Z_{2, \text{exact}}^{(3)}}$ obtained from performing Wick contractions for Eq.\ \eqref{Z_2_3_product_Mijk}. 
We report the expression below:

\begin{widetext}
    \begin{align}
    \average{Z_{2, \text{exact}}^{(3)}} = \frac{1}{\mathcal{N}^4} 
    \Bigg\{& \sum_{E^A_1, E^A_2} \sum_{E^B_1, E^B_2} e^{\sum_{i=1}^2 S_A(E^A_i) + \sum_{j=1}^2 S_B(E^B_j) + \sum_{i, j=1}^2 S_C(E - E^A_i - E^B_j)} + (A \leftrightarrow B \leftrightarrow C)\\ \nonumber
    + & 2\sum_{E^A_1}\sum_{E^B_1}\sum_{E^C_1} e^{S_A(E^A_1) + S_B(E^B_1) + S_C(E^C_1) + 2S_A(E - E^B_1 - E^C_1) + 2S_B(E - E^A_1 - E^C_1)} + (A \leftrightarrow B \leftrightarrow C) \\ \nonumber 
    + & 2\sum_{E^A_1}\sum_{E^B_1} e^{S_A(E^A_1) + S_B(E^B_1) + 3S_C(E - E^A_1 - E^B_1)} + (A \leftrightarrow B \leftrightarrow C)\\ \nonumber 
    + & 9\sum_{E^A_1}\sum_{E^B_1} e^{2S_A(E^A_1) + 2S_B(E^B_1) + 2S_C(E - E^A_1 - E^B_1)}
    \Bigg\}. 
\end{align}
\end{widetext}

Here, since $\average{Z_{2, \text{exact}}^{(3)}}$ is symmetric in subsystems $A$, $B$ and $C$, each term in the first three lines has two additional counterparts generated by permuting subsystem labels.
For clarity of presentation, the above expression assumes that the window function $F(\Delta)$ is a delta function. 
In the full analytical calculation for comparison with numerics, we instead allow a finite energy window of width $\Delta E$.
This yields two slightly different types of terms that both reduce to the form $\sum_{E^A_1}\sum_{E^B_1} e^{2S_A(E^A_1) + 2S_B(E^B_1) + 2S_C(E - E^A_1 - E^B_1)}$ in the $\Delta E \to 0$ limit.
Both types of terms involve two energies $E^K_1$ and $E^K_2$ per subsystem, but eight terms require that the four sums $E^A_1 + E^B_1 + E^C_2$, $E^A_1 + E^B_1 + E^C_1$, $E^A_2 + E^B_1 + E^C_1$, $E^A_1 + E^B_2 + E^C_1$ are all in the range $[E - \Delta E, E + \Delta E]$, while one other term instead imposes this total energy constraint on $E^A_1 + E^B_1 + E^C_1$, $E^A_2 + E^B_1 + E^C_2$, $E^A_1 + E^B_2 + E^C_2$, $E^A_2 + E^B_2 + E^C_1$.

\subsection{Dependence on the energy-window width}
\label{appendix:check_on_delta_E}

In the analytical calculation of $\average{S_2^{(3)}}$ in Sec.\ \ref{sec:exact_diagonalization_multi_entropy}, we used an energy window function with width $\Delta E = 0.7$. 
Intuitively, following similar arguments as for the bipartite case \cite{Lu_2017, Murthy_2019}, as long as $\Delta E$ is $O(1)$ and much greater than individual level spacing, small changes in $\Delta E$ should not affect the analytical expression in any significant way.

To verify that $\average{S_2^{(3)}}$ is insensitive to the width $\Delta E$ of the energy window function, we evaluate the analytical predictions of $\average{S_2^{(3)}}$ using a range of window widths including $\Delta E = 0.7$. 
We perform this check for the mixed-field Ising model (Eq.\ \eqref{mixed_field_Ising_model}) with four sets of parameters, all with the same longitudinal field $h = 0.35$, and different transverse fields $g = 1.0, 1.1, 1.2$ and $1.3$ respectively. 
These parameters are chosen around the maximally chaotic regime. For each set of parameters, we examine two different energies, one relatively close to the center of spectrum at $\beta = 0.05$, and one farther from the center at $\beta = 0.2$ (the exact diagonalization results for R\'enyi-2 multi-entropy are still consistent with the analytical prediction at this energy).
The density of states used in the analytical predictions is obtained from exact diagonalization of the corresponding subsystem Hamiltonians at these parameters. 
We choose subsystem sizes $V_A = V_B = V_C = V/3$.
For ease of comparison, we use the value of $\average{S_2^{(3)}}$ computed at $\Delta E = 0.7$ as the baseline, and normalize all results by this baseline value.

\begin{figure}
     \centering
        \includegraphics[width=0.49\textwidth]{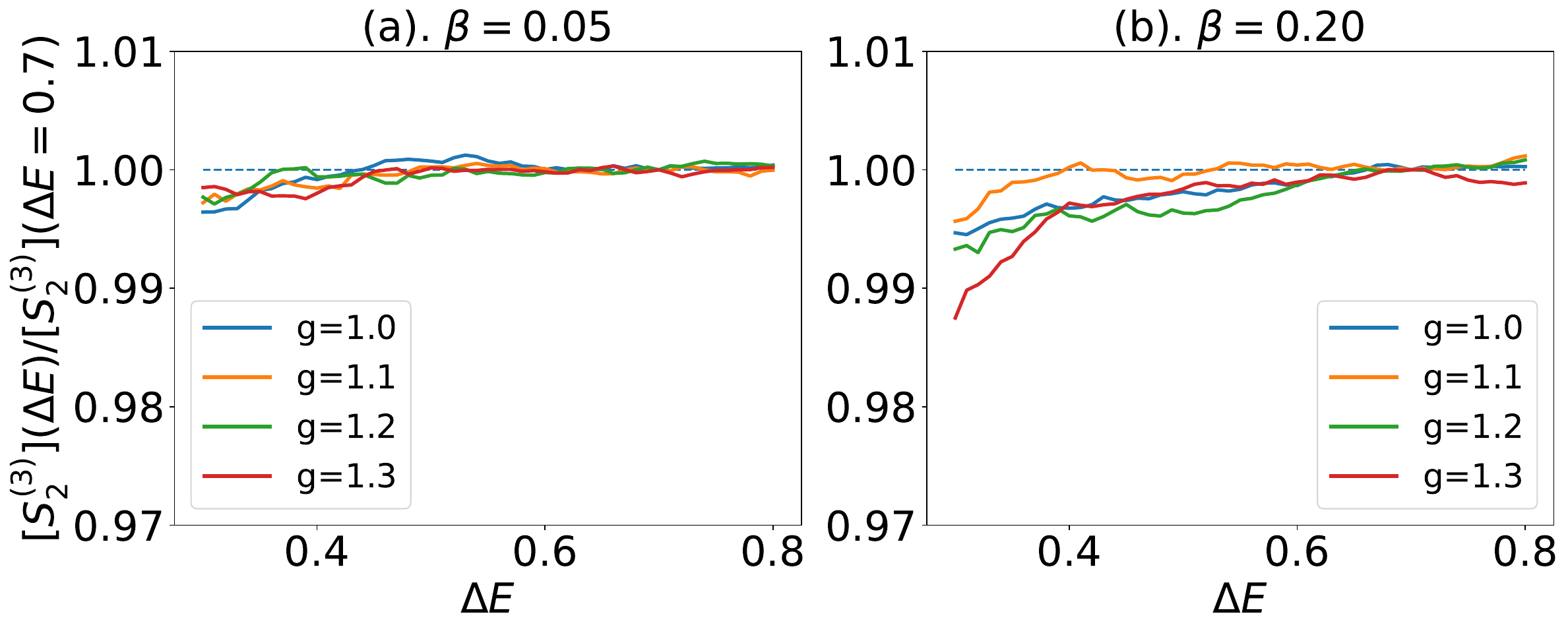}
    \caption{Ratio between the analytical prediction for $\average{S_2^{(3)}}$ and the baseline at different energy-window width $\Delta E$. We plot the ratio $\average{S_2^{(3)}}(\Delta E)/\average{S_2^{(3)}}(\Delta E=0.7)$ for $h=0.35$ and the indicated values of $g$ 
    at equal subsystem sizes $V_A=V_B=V_C=V/3$ and energies corresponding to (left) $\beta = 0.05$ and (right) $\beta = 0.2$. 
    }
    \label{fig:deltaE_check}
\end{figure}

Fig.~\ref{fig:deltaE_check} shows the ratio between $S_2^{(3)}$ obtained at different values of $\Delta E$ to the baseline.
Across all values of $g$ and for both inverse temperatures considered, the ratio remains close to unity over the entire range of $\Delta E$, with deviations at the percent level or below.
This demonstrates that $S_2^{(3)}$ is insensitive to the precise choice of energy-window width within a physically reasonable range. 

\section{More on the Properties of $\rho_{AA^*}^{(2)}$ and $\rho_{AA^*}$}

\subsection{$\rho_{AA^*}^{(2)}$ and $\rho_{AA^*}$ when $V_C \gg V_A + V_B$}
\label{appendix:rho_AA_from_rho_AB}

It has been shown in the tripartite case that when $V_C \gg V_A + V_B$, the individual reduced density matrix $\rho_{AB}$ is well-described by a semi-classical form \cite{Dymarsky_2018} consistent with the ``ergodic bipartition" ansatz \cite{Lu_2017}.
Specifically, assuming that $F(\Delta)$ in Eq.\ \eqref{ergodic_bipartition} is a delta function peaked at $\Delta = 0$, the leading order contributions to individual $\rho_{AB}$ matrices take the form:
\begin{equation}
    \rho_{AB} = e^{-S(E)} \sum_{i, j} e^{S_C(E - E_i - E_j)} |E_i\rangle_A |E_j\rangle_B \langle E_i|_A \langle E_j|_B,
    \label{average_rho_AB_analytical}
\end{equation}
which is a diagonal matrix. 
The off-diagonal terms are conjectured and shown numerically to be exponentially suppressed \cite{Dymarsky_2018}.

\begin{widetext}
Eq.\ \eqref{average_rho_AB_analytical} leads to canonical purifications $|\sqrt{\rho_{AB}}\rangle$ and $|\rho_{AB}^{(2)}\rangle$ of the following form:
\begin{align}
    |\sqrt{\rho_{AB}}\rangle & = e^{-S(E)/2} 
    \sum_{i, j} e^{S_C(E - E_i - E_j)/2} |E_i\rangle_A |E_j\rangle_B |E_i\rangle_{A^*} |E_j\rangle_{B^*}, \\
    |\rho_{AB}^{(2)}\rangle & = \frac{e^{-S(E)}}{\sqrt{\text{tr}(\rho_{AB}^2)}} 
    \sum_{i, j} e^{S_C(E - E_i - E_j)} |E_i\rangle_A |E_j\rangle_B |E_i\rangle_{A^*} |E_j\rangle_{B^*}.
\end{align}
And the corresponding reduced density matrices $\rho_{AA^*}$ and $\rho_{AA^*}^{(2)}$ can be expressed as
\begin{align}
     & \rho_{AA^*} = e^{-S(E)} \sum_{i_1, i_2} \sum_j e^{S_B(E_j)} e^{S_C(E - E_{i_1} - E_j)/2} e^{S_C(E - E_{i_2} - E_j)/2} |E_{i_1}\rangle_A |E_{i_1}\rangle_{A^*} \langle E_{i_2}|_A \langle E_{i_2}|_{A^*}, 
     \label{average_rho_AA_from_rho_AB} \\
    & \rho_{AA^*}^{(2)} = \frac{e^{-2S(E)}}{\text{tr}(\rho_{AB}^2)} \sum_{i_1, i_2} \sum_j e^{S_B(E_j)} e^{S_C(E - E_{i_1} - E_j)} e^{S_C(E - E_{i_2} - E_j)} |E_{i_1}\rangle_A |E_{i_1}\rangle_{A^*} \langle E_{i_2}|_A \langle E_{i_2}|_{A^*}.
    \label{average_rho_AA_2_from_rho_AB}
\end{align}
It is clear that Eq.\ \eqref{average_rho_AA_2_from_rho_AB}, derived from the universal form of $\rho_{AB}$, is consistent with Eq.\ \eqref{eq:average_rho_AA_2_analytical} derived from Eq.\ \eqref{ergodic_tripartition}.

\subsection{$\rho_{AA^*}^{(2)}$ in the $V_A/V \to 0$ limit}
\label{appendix:vanishing_VA_rho_AA_2_calculation}

In this Appendix, we calculate $\rho_{AA^*}^{(2)}$ in the three regimes $L_B \gg L_C$, $L_B \ll L_C$ and $L_B \approx L_C$, in the thermodynamic limit where $V \to \infty$ while size $V_A$ of subsystem $A$ and energy density $E/V$ are held finite and fixed. 

First, we note that the normalization
\begin{align}
\text{tr}(\rho_{AB}^2)
= e^{-2S(E)}\sum_{E_i^A, E_j^B} (e^{S_{AB}(E_i^A + E_j^B) + 2S_C(E - E_i^A - E_j^B)}
+ e^{2S_{AB}(E_i^A + E_j^B) + S_C(E - E_i^A - E_j^B)}).
\end{align}

When $L_B \gg L_C$, the diagonal terms dominate. We first perform the saddle point approximation on the diagonal term, and write
\begin{align}
    & [\rho_{AA^*}^{(2)}]_{i_1 V_A + i_1', i_1 V_A + i_1'} =
     \frac{e^{-2S(E)}}{\text{tr}(\rho_{AB}^2)} e^{S_C(E_{k_1}^{C*}) + S_B(E - E_{i_1}^A - E_{k_1}^{C*})+ S_B(E - E_{i_1'}^A - E_{k_1}^{C*})},
\end{align}
where $E_{k_1}^{C*}$ is the value of $E_{k_1}^C$ that maximizes $S_C(E_{k_1}^C) + S_B(E - E_{i_1}^A - E_{k_1}^C) + S_B(E - E_{i_1'}^A - E_{k_1}^C)$. 
Since $V_A/V \to 0$, we further approximate $S_B(E - E_{i_1}^A - E_{k_1}^{C*}) \approx S_B(E - E_{k_1}^{C*}) - \beta_B E_{i_1}^A$, so 
\begin{align}
    & [\rho_{AA^*}^{(2)}]_{i_1 V_A + i_1', i_1 V_A + i_1'}
     = \frac{e^{-2S(E)}e^{S_C(E_{k_1}^{C*}) + 2S_B(E - E_{k_1}^{C*})}}{\text{tr}(\rho_{AB}^2)}  e^{-\beta_B(E_{i_1}^A + E_{i_1'}^A)},
\end{align}
where $\beta_B$ is the inverse temperature in subsystem $B$ at energy $E - E_{k_1}^{C*}$. Since $E_{i_1}^A, E_{i_1'}^A \sim O(V_A) \sim O(1)$ while $E_{k_1}^C, E \sim O(V)$, we can further assume that the change in $E_{k_1}^{C*}$ due to changes in $E_{i_1}^A, E_{i_1'}^A$ is small compared to saddle width and therefore does not affect the leading saddle contribution $S_C(E_{k_1}^{C*}) + 2S_B(E - E_{k_1}^{C*})$. Then, we can write
\begin{equation}
    \rho_{AA^*}^{(2)} \approx \frac{e^{S_C(E_{k_1}^{C*}) + 2S_B(E - E_{k_1}^{C*})}}{\sum_{E_{k}}e^{S_C(E_k)+2S_{AB}(E-E_k)}}  e^{-\beta_B(H_A + H_{A^*})}.
\end{equation}

In the limit $L_C \gg L_B$, the block terms dominate. We can similarly make the saddle point approximation for the block term
\begin{align}
     [\rho_{AA^*}^{(2)}]_{i_1V_A+i_1, i_2V_A+i_2}
     = \frac{e^{-2S(E)}}{\text{tr}(\rho_{AB}^2)} e^{S_B(E_{j_1}^{B*}) + S_C(E - E_{i_1}^A - E_{j_1}^{B*}) + S_C(E - E_{i_2}^A - E_{j_1}^{B*})},
\end{align}
where $E_{j_1}^{B*}$ is the value of $E_{j_1}^B$ that maximizes $S_B(E_{j_1}^B) + S_C(E - E_{i_1}^A - E_{j_1}^B) + S_C(E - E_{i_2}^A - E_{j_1}^B)$. We approximate $S_C(E - E_{i_1}^A - E_{j_1}^{B*}) \approx S_C(E - E_{j_1}^{B*}) - \beta_C E_{i_1}^A$.
In this case, instead of looking for an entanglement Hamiltonian for $\rho_{AA^*}^{(2)}$, we can consider the purification $|\tilde{\rho}_A^{(2)}\rangle$ of a thermal reduced density matrix $\tilde{\rho}_A = e^{-\beta_C H_A}$, which takes the form $|\tilde{\rho}_A^{(2)}\rangle = \sum_i e^{-\beta_C E_i^A}|E_i^A\rangle_A|E_i^A\rangle_{A^*}$. 
Since $E_i^A \sim O(1)$ while $E_j^B, E \sim O(V)$, the changes in leading saddle contribution across different $E_{i_1}^A$ and $E_{i_2}^A$ are negligible. Thus, $\rho_{AA^*}^{(2)} \propto |\tilde{\rho}_A^{(2)} \rangle\langle\tilde{\rho}_A^{(2)}|$. Specifically, we can write
\begin{equation}
    \rho_{AA^*}^{(2)} \approx \frac{e^{S_B(E_{j_1}^{B*}) + 2S_C(E - E_{j_1}^{B*})}}{\sum_{E_k} e^{2S_C(E_k) + S_{AB}(E-E_k)}} |\tilde{\rho}_A^{(2)}\rangle\langle\tilde{\rho}_A^{(2)}|.
\end{equation}

Finally, we note that in the previous calculations for the diagonal and block terms, we didn't make assumptions on the relative sizes for subsystems $B$ and $C$. Thus, in the $L_B \approx L_C$ regime where both the block and diagonal terms are present, $\rho_{AA^*}^{(2)}$ combines the forms in the $L_B \gg L_C$ and $L_B \ll L_C$ regimes:
\begin{align}
    \rho_{AA^*}^{(2)} &\approx  \frac{e^{-2S(E)}}{\text{tr}(\rho_{AB}^2)}
    \Big\{e^{S_C(E_{k_1}^{C*}) + 2S_B(E - E_{k_1}^{C*})}e^{-\beta_B (H_A + H_{A^*})}
    + e^{S_B(E_{j_1}^{B*}) + 2S_C(E - E_{j_1}^{B*})}|\tilde{\rho}_A^{(2)}\rangle\langle\tilde{\rho}_A^{(2)}
    |\Big\}.
\end{align}
\end{widetext}

\bibliography{references}

\end{document}